\documentclass[aps,prl,twocolumn,preprintnumbers,superscriptaddress,dblfloatfix,nofootinbib]{revtex4-1}
\usepackage[utf8]{inputenc}
\usepackage{lmodern}
\usepackage{amsmath,amssymb,mhchem}
\usepackage{graphicx}
\usepackage{url}
\usepackage{color}
\usepackage{enumitem}
\usepackage{subfigure}
\usepackage[dvipsnames]{xcolor}
\usepackage[colorlinks=true,breaklinks=true]{hyperref}
\hypersetup{allcolors=[rgb]{0.0 0.0 0.6},linkcolor=[rgb]{0.75 0.05 0.05}}
\usepackage{bm}
\usepackage{epsfig}
\usepackage{amsmath}
\usepackage{amssymb}
\usepackage{slashed}
\usepackage{color}
\usepackage{accents}
\usepackage[dvipsnames]{xcolor}
\usepackage[colorlinks=true,breaklinks=true]{hyperref}
\hypersetup{allcolors=[rgb]{0.0 0.0 0.6},linkcolor=[rgb]{0.75 0.05 0.05}}

\newcommand{\exclude}[1]{}

\begin{document}

\preprint{IPMU20-0105}
\preprint{YITP-20-125}

\title{Testing Stochastic Gravitational Wave Signals from Primordial Black Holes \\ with Optical Telescopes
}

\author{Sunao Sugiyama}
\affiliation{Kavli Institute for the Physics and Mathematics of the Universe (WPI), UTIAS \\The University of Tokyo, Kashiwa, Chiba 277-8583, Japan}
\affiliation{Department of Physics, The University of Tokyo, 7-3-1 Hongo, Bunkyo-ku, Tokyo 113-0033 Japan}

\author{Volodymyr Takhistov}
\affiliation{Kavli Institute for the Physics and Mathematics of the Universe (WPI), UTIAS \\The University of Tokyo, Kashiwa, Chiba 277-8583, Japan}
\affiliation{Department of Physics and Astronomy, University of California, Los Angeles \\ Los Angeles, California, 90095-1547, USA}

\author{Edoardo Vitagliano}
\affiliation{Department of Physics and Astronomy, University of California, Los Angeles \\ Los Angeles, California, 90095-1547, USA}

\author{Alexander Kusenko}
\affiliation{Department of Physics and Astronomy, University of California, Los Angeles \\ Los Angeles, California, 90095-1547, USA}
\affiliation{Kavli Institute for the Physics and Mathematics of the Universe (WPI), UTIAS \\The University of Tokyo, Kashiwa, Chiba 277-8583, Japan}

\author{Misao Sasaki}
\affiliation{Kavli Institute for the Physics and Mathematics of the Universe (WPI), UTIAS \\The University of Tokyo, Kashiwa, Chiba 277-8583, Japan}
\affiliation{Center for Gravitational Physics, Yukawa Institute for
Theoretical Physics, \\
Kyoto University, Kyoto 606-8502, Japan}
\affiliation{Leung Center for Cosmology and Particle Astrophysics,
National Taiwan University, 
Taipei 10617, Taiwan}

\author{Masahiro Takada}
\affiliation{Kavli Institute for the Physics and Mathematics of the Universe (WPI), UTIAS \\The University of Tokyo, Kashiwa, Chiba 277-8583, Japan}

\date{\today}

\begin{abstract}
Primordial black holes (PBHs) can constitute the predominant fraction of dark matter (DM) if PBHs reside in the currently unconstrained ``sublunar'' mass range. PBHs originating from scalar perturbations generated during inflation can naturally appear with a broad spectrum in a class of models. The resulting stochastic gravitational wave (GW) background generated from such PBH production can account for the recently reported North American Nanohertz Observatory for Gravitational Waves (NANOGrav) pulsar timing array data signal, and will be testable in future GW observations by interferometer-type experiments such as Laser Interferometer Space Antenna (LISA). We show that the broad mass function of such PBH~DM is already being probed by Subaru Hyper Suprime-Cam (HSC) microlensing data and is consistent with a detected candidate event. Upcoming observations of HSC will be able to provide an independent definitive test of the stochastic GW signals originating from such PBH~DM production scenarios.
\end{abstract}
\maketitle


Primordial black holes (PBHs), formed in the early Universe prior to any galaxies and stars, are a viable dark matter (DM) candidate~(e.g.~\cite{Zeldovich:1967,Hawking:1971ei,Carr:1974nx,GarciaBellido:1996qt,Green:2000he,Khlopov:2008qy,Frampton:2010sw,Kawasaki:2016pql,Cotner:2016cvr,Cotner:2017tir,Green:2016xgy,Carr:2016drx,Inomata:2016rbd,Pi:2017gih,Inomata:2017okj,Garcia-Bellido:2017aan,Inoue:2017csr,Georg:2017mqk,Inomata:2017bwi,Kocsis:2017yty,Ando:2017veq,Cotner:2016cvr,Cotner:2017tir,Germani:2017bcs,Nishikawa:2017chy,Germani:2018jgr,Cotner:2018vug,Sasaki:2018dmp,Carr:2018rid,Cotner:2019ykd,Germani:2019zez,Inman:2019wvr,Flores:2020drq,Deng:2017uwc,Green:2020jor,Kusenko:2020pcg,deFreitasPacheco:2020wdg}). PBHs could also play a central role in a variety of astrophysical phenomena, such as progenitors~\cite{Nakamura:1997sm,Clesse:2015wea,Bird:2016dcv,Raidal:2017mfl,Eroshenko:2016hmn,Sasaki:2016jop,Clesse:2016ajp,Fuller:2017uyd}
for the Laser Interferometer Gravitational-Wave Observatory (LIGO) gravitational wave (GW) events~\cite{Abbott:2016blz,Abbott:2016nmj,Abbott:2017vtc}, seeds for formation of super-massive black holes~\cite{Bean:2002kx,Kawasaki:2012kn,Clesse:2015wea}, as well as the source of new signals~\cite{Fuller:2017uyd,Takhistov:2017nmt,Takhistov:2017bpt} from compact star disruptions due to PBH capture, among others. Depending on the formation time, the resulting PBHs can span many decades of orders of magnitude in mass. PBHs
formed with mass above the Hawking evaporation limit of $\sim 10^{15}$~g survive until present and contribute to the DM abundance. 
PBHs residing in the smaller ``sublunar'' mass range of $\sim 10^{-16} - 10^{-10} M_{\odot}$ can constitute the entirety of DM~\cite{Katz:2018zrn,Smyth:2019whb,Montero-Camacho:2019jte}.

Recently, the North American Nanohertz Observatory for Gravitational Waves (NANOGrav) collaboration has reported a putative signal from a 12.5 years analysis of pulsar timing array data~\cite{Arzoumanian:2020vkk}. The signal is consistent with a stochastic GW background (SGWB) with a GW strain amplitude of $A \sim 10^{-15}$ and a frequency of $f \sim 10^{-9}$ Hz, resulting in a radiation-dominated post-inflationary Universe SGWB of $\Omega_{\rm GW} h^2 \sim 10^{-10}$ with $h$ being the dimensionless Hubble parameter. While the reported NANOGrav signal is in tension with their own previous 11 years data analysis~\cite{Arzoumanian:2018saf}, as well as with the Parkes Pulsar Timing Array (PPTA)~\cite{Shannon:2015ect} analysis, it has been argued that previous pulsar timing array results have been over-interpreted in astrophysical terms, leading to suppressed potential signals. In the nHz frequency range, a significant source of GWs is expected to originate from supermassive black hole binaries  (e.g.~\cite{Sesana:2004sp,Burke-Spolaor:2018bvk}). However, predictions of this SGWB contribution suffer from significant uncertainties. The reported signal has been also interpreted in terms of cosmic strings~\cite{Blasi:2020mfx,Ellis:2020ena,Buchmuller:2020lbh,Samanta:2020cdk}, domain walls~\cite{Bian:2020bps}, phase transitions~\cite{Nakai:2020oit,Addazi:2020zcj,Ratzinger:2020koh, Li:2020cjj}, and tensor perturbations from inflation~\cite{Vagnozzi:2020gtf}.

During PBH production in the early Universe from enhanced curvature perturbations, a SGWB will also be generated at second-order. This allows to naturally associate the reported NANOGrav observations with PBH formation~\cite{Vaskonen:2020lbd,DeLuca:2020agl,Kohri:2020qqd}. 
 The studies in the literature do this in a distinct manner and differ in their mass-range for the resulting dominant PBH component. Particularly intriguing is the possibility of a broad PBH mass function~\cite{DeLuca:2020ioi} arising in a class of motivated models~\cite{Wands:1998yp,Leach:2000yw,Leach:2001zf,Biagetti:2018pjj} that can account for the entirety of DM in the open lower-mass parameter space window as well as SGWB signal simultaneously~\cite{DeLuca:2020agl}.
Following Ref. \cite{DeLuca:2020agl}, we employ the broad flat perturbation power spectrum that allows to simultaneously relate the NANOGrav signal to solar-mass PBHs as well as account for small asteroid-mass $\sim 10^{-12} M_{\odot}$ PBHs that constitute all of the DM. In contrast, Ref.~\cite{Vaskonen:2020lbd,Kohri:2020qqd} consider variations of log-normal power-spectrum and do not address DM abundance with PBHs. The   studies differ in their assumptions for PBH abundance calculations, such as the choice of the window function.  

In this work, we demonstrate that a broad PBH mass function, responsible for DM and that simultaneously can account for the reported SGWB signal, is already being probed by Subaru Hyper Suprime-Cam (HSC), and will be definitively tested with upcoming observations of HSC and other optical telescopes. In contrast to other measurements designed to probe the SGWB, such as those of the future Laser Interferometer Space Antenna (LISA), this constitutes an independent test of the reported signal. The potential of HSC to probe PBH models with a broad mass function has been recently stressed in the context of other general PBH formation scenarios~\cite{Kusenko:2020pcg}.

We now briefly describe PBH and SGWB production from a broad mass function that could arise in a class of models such as~\cite{Wands:1998yp,Leach:2000yw,Leach:2001zf,Biagetti:2018pjj}, following Ref.~\cite{DeLuca:2020agl}. We consider the
broad and flat power spectrum of curvature perturbations of the form~\cite{DeLuca:2020ioi}
\begin{equation} \label{eq:spec}
    \mathcal{P}_{\zeta}(k) \simeq A_\zeta  \Theta(k_s - k) \Theta(k - k_l) ,
\end{equation}
where $k_s \gg k_l$, $\Theta$ is the Heaviside step function and $A_\zeta$ is the amplitude. This also gives the spectrum of overdensity $\delta$ perturbations $\mathcal{P}_{\delta}(k)$.

\begin{figure*}[t]
\begin{center}
\includegraphics[width=1\columnwidth]{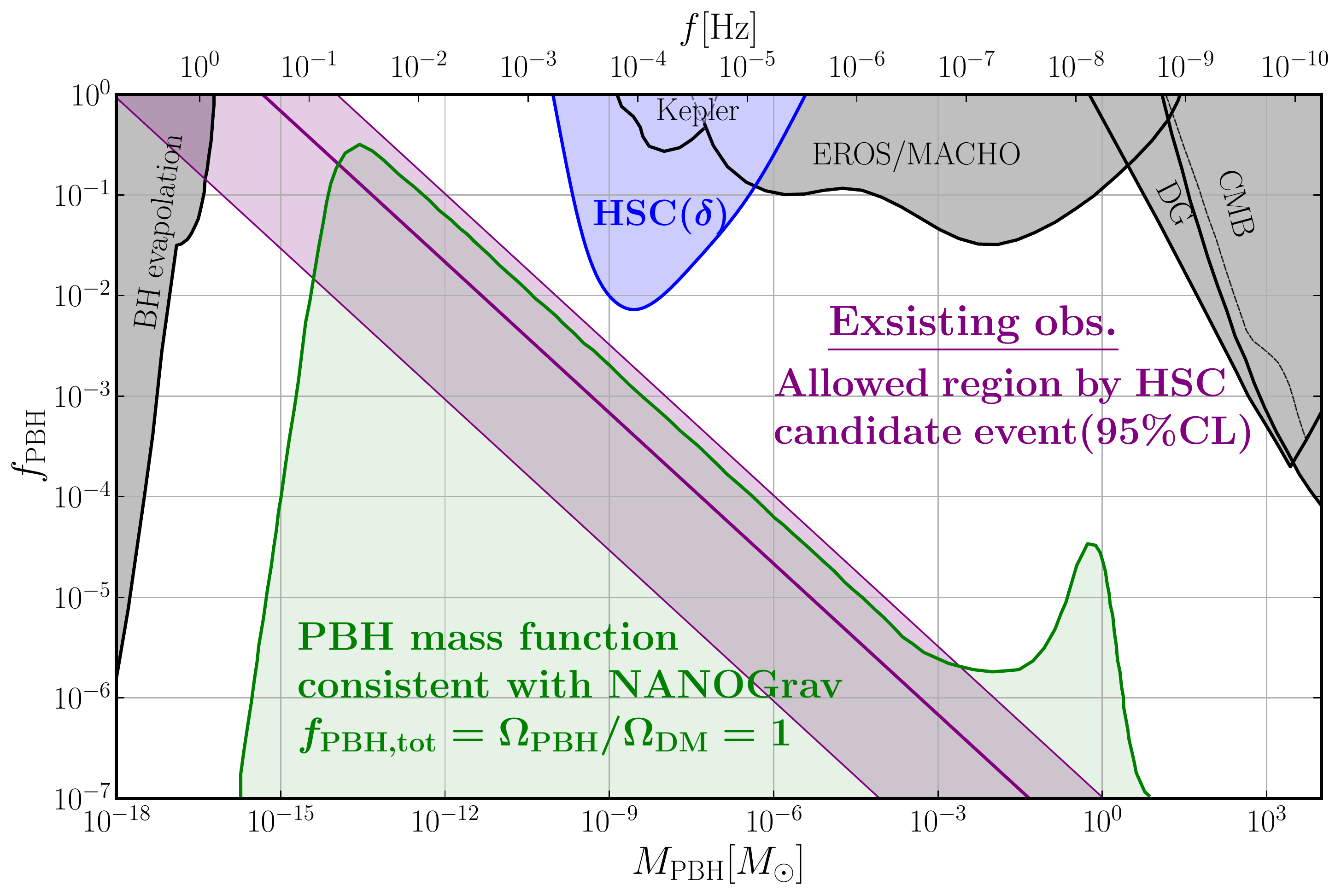} 
\includegraphics[width=1\columnwidth]{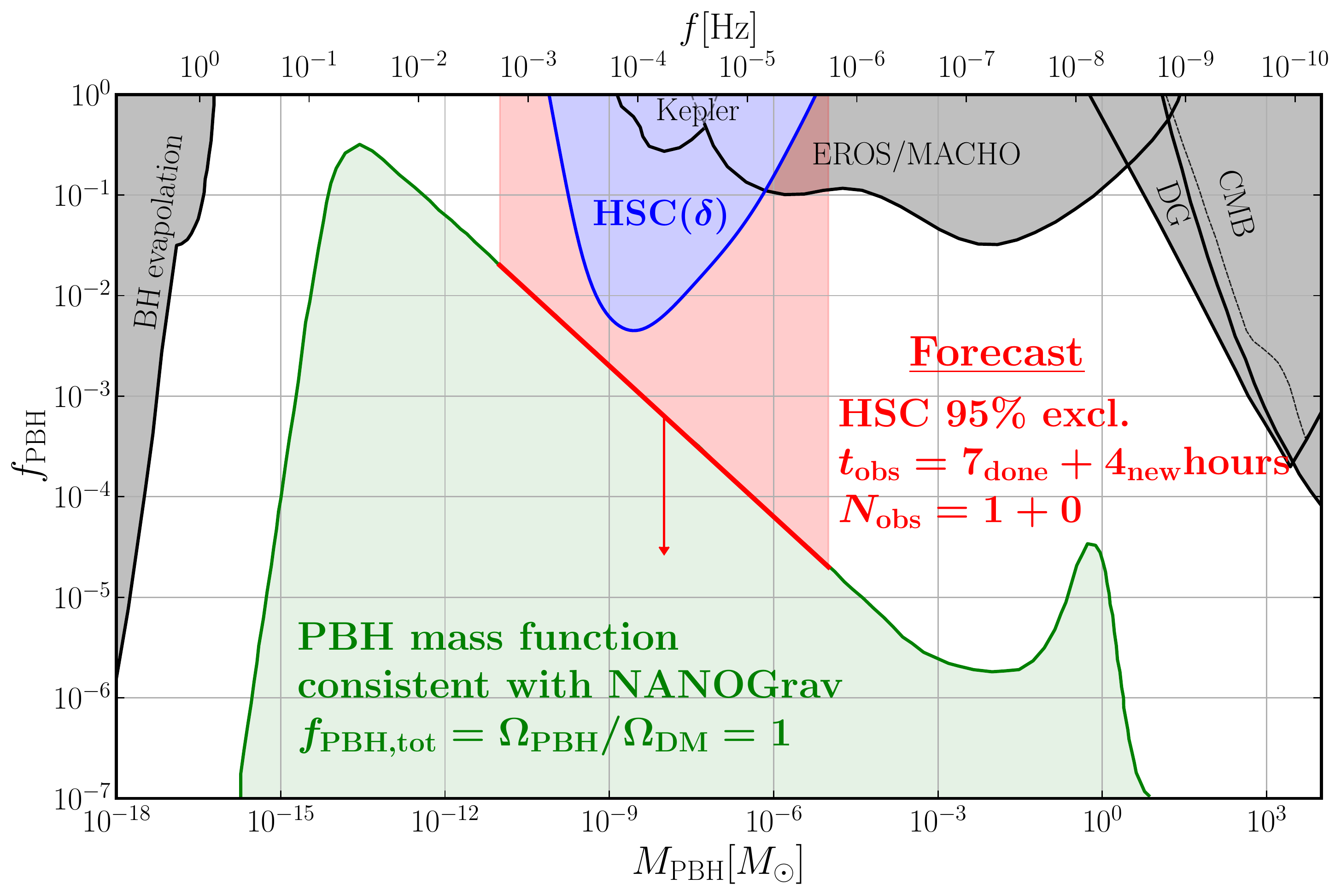} 
\caption{
[Left] Allowed normalization range (shown in purple) for a PBH mass function $\propto M_{\rm PBH}^{-1/2}$ consistent with the HSC candidate event reported after 7 hours of observations~\cite{Niikura:2017zjd}. The thick purple line denotes the best fit normalization. The HSC constraint on the total PBH DM fraction $f_{\rm PBH, tot}$, assuming a monochromatic mass function, is also shown in blue. We account for the finite source effects~\cite{Smyth:2019whb,Sugiyama:2019dgt}. The green line shows the mass function which is consistent with the reported NANOGrav SGWB signal and normalized to be $f_{\rm PBH,tot}=\Omega_{\rm PBH}/\Omega_{\rm DM}=1$ \cite{DeLuca:2020agl}.
[Right] Forecast of the exclusion region for the normalization of a PBH mass function $\propto M_{\rm PBH}^{-1/2}$ spectrum (shown in red), combining existing 7 hours of observation and additional 4 hours of observation (assuming 1 and 0 event for each, respectively). The blue region shows the constraint on the total PBH DM fraction $f_{\rm PBH,tot}$, assuming a monochromatic mass function, obtained with the same observation time ($4+7=11$ hours in total).
Constraints from  extragalactic $\gamma$-rays from BH evaporation~\cite{Carr:2020gox} (additional constraints in this region have been also recently suggested~\cite{Arbey:2019vqx,Dasgupta:2019cae,Laha:2019ssq,DeRocco:2019fjq,Laha:2020vhg,Laha:2020ivk}), microlensing Kepler data~\cite{Griest:2013aaa}, MACHO/EROS/OGLE microlensing~\cite{Tisserand:2006zx}, the accretion effects on the CMB observables~\cite{Ali-Haimoud:2016mbv,Poulin:2017bwe,Poulter:2019ooo} as well as dwarf galaxy heating \cite{Lu:2020bmd} are also displayed.
}
\label{fig:PBHfull}
\end{center}
\end{figure*}

PBHs form when a sufficient overdense region, corresponding to the density
fluctuations with a sufficiently large amplitude at a certain scale, enters the Hubble horizon. Hence, 
given the perturbation spectrum, the energy density in PBHs at formation can be found from Press-Schechter formalism as~\cite{Sasaki:2018dmp} 
\begin{equation}
    \beta(M_{\rm PBH}) = \int_{\delta_c} d\delta \dfrac{M_{\rm PBH}}{M_H}P(\delta,\sigma_{\rm PBH}),
\end{equation}  
where $\delta_c$ is the critical density contrast threshold for collapse,
and near the collapse threshold one has to take into account the scaling of the PBH mass with respect to the horizon mass\footnote{The horizon mass
is related to the characteristic comoving frequency of perturbation as
$M_H \simeq 33 (10^{-9}\textrm{Hz}/f)^2 M_{\odot}$.} $M_{\rm PBH}/M_H = \kappa (\delta - \delta_c)^{\gamma_c}$  with $\kappa \simeq 3.3$  and $\gamma_c \simeq 0.36$~\cite{Choptuik:1993,Evans:1994pj,Niemeyer:1997mt,Niemeyer:1999ak,Koike:1995jm}.
Here, $P(\delta,\sigma_{\rm PBH})$ is the probability distribution of density fluctuations entering the horizon, assumed to be Gaussian with a variance
\begin{equation}
    \sigma_{\rm PBH}^2 = \int d{\rm ln}k \, W^2 (k, R_H) T^2 (k, R_H)\mathcal{P}_\delta(k),
\end{equation}
where $W(k,R_H)$ is a window function for smoothing over the horizon scale $R_H \sim 1/aH$, and $T(k,R_H)$ is a transfer function smoothing over sub-horizon modes~\cite{DeLuca:2020ioi}. After matter-radiation equality the DM fraction consisting of PBHs can be expressed as a mass function (e.g.~\cite{Byrnes:2018clq})
\begin{equation}\label{eq:mass_function}
    f_{\rm PBH}(M_{\rm PBH}) = \dfrac{1}{\Omega_{\rm DM}} \Big(\dfrac{M_{\rm eq}}{M_{\rm PBH}}\Big)^{1/2}\beta(M_{\rm PBH}) ,
\end{equation}
where $M_{\rm eq} \simeq 3 \times 10^{17} M_{\odot}$ is the horizon mass at equality, and $\Omega_{\rm DM}$ is cold dark matter density. The total contribution of PBHs to DM $f_{\rm PBH, tot}$ is found by integrating $f_{\rm PBH}$ over ${\rm ln} \, M_{\rm PBH}$.

Enhanced curvature perturbations, responsible for PBH generation, also induce GWs at second order (see e.g. Ref.~\cite{Cai:2018dig,Ananda:2006af}). The contribution today of this SGWB  is given by~\cite{Kohri:2018awv,Espinosa:2018eve,DeLuca:2020agl}
\begin{align} \label{eq:sgwb}
    \Omega_{\rm GW}(f) =&  \dfrac{c_g \Omega_{r,0}}{972} \int \int_S dx dy \dfrac{x^2}{y^2} \Big[1 - \dfrac{(1+x^2-y^2)^2}{4x^2}\Big]^2 \notag\\
    & \times \mathcal{P}_{\zeta}(kx) \mathcal{P}_{\zeta}(ky) \mathcal{I}^2(x,y) ,
\end{align}
where $k = 2 \pi f$, $\Omega_{r,0}$ is the radiation abundance today, $\mathcal{I}(x,y)$ is the kernel function, and the integration region $S$ covers $x > 0$ and $|1-x| \leq y \leq 1 + x$. The factor $c_g$ accounts for the change in number of degrees of freedom of thermal radiation during time evolution.

In Ref.~\cite{DeLuca:2020agl} it was shown, by fitting Eq.~\eqref{eq:sgwb}, that the reported NANOGrav signal can result from the  perturbation spectrum of Eq.~\eqref{eq:spec}, peaking at $M_{\rm PBH} \sim 10^{-14} M_{\odot}$ and responsible for PBH~DM, with $A_{\zeta} \simeq 5.8 \times 10^{-3}$, $k_s = 10^9 k_l \simeq 1.6$ Hz, and $f_{\rm PBH, tot} = 1$. We display this result in Fig. \ref{fig:PBHfull}. The peak of the
 mass function denotes the horizon mass when the shortest scale $\sim 1/k_s$ re-enters. The sub-dominant peak around solar-mass results from a change in the equation of state at the QCD phase transition\footnote{We note that the softening of the equation of state due to the QCD phase transition may also affect the spectrum of second order GWs~\cite{Hajkarim:2019nbx}. We do not expect the effect to be very significant.}~\cite{Byrnes:2018clq}. The PBH mass function between peaks follows $\sim M^{-1/2}$ functional dependence\footnote{We note that with a different definition (e.g. Ref.~\cite{DeLuca:2020ioi}), the mass function scales as $\sim M^{-3/2}$.}. Since the SGWB spectrum from PBHs extends over many orders of magnitude, this scenario can be tested with future GW observations such as those of LISA.

Here we suggest that optical telescopes can provide an independent test of this explanation for the NANOGrav signal and will provide a definitive probe in the near future. Even though the mass function of PBHs constituting DM peaks at much smaller masses, where microlensing effects are negligible, the broad mass function has a tail overlapping with the HSC sensitivity range. This is actually consistent with HSC observations of Andromeda galaxy (M31)~\cite{Niikura:2017zjd}, which reported a candidate event consistent with PBH at $f_{\rm PBH}(M \sim 10^{-9} M_{\odot}) \sim 10^{-2}$. While these observations lend credibility to the hypothesis of PBHs being the source of the SGWB detected by NANOGrav, additional HSC observations will be able to test this scenario.

Following Ref.~\cite{Kusenko:2020pcg}, we estimate the reach of HSC, exploiting results from HSC Monte Carlo simulations as well as their analysis tools~\cite{Niikura:2017zjd}. The expected number of observed microlensing events reads
\begin{widetext}
\begin{equation}
N_{\rm exp} \Big(\dfrac{\Omega_{\rm PBH}}{\Omega_{\rm DM}}\Big) = \dfrac{\Omega_{\rm PBH}}{\Omega_{\rm DM}} \int\! dM \int_0^{t_{\rm obs}}\! \dfrac{\mathrm{d}t_{\rm FWHM}}{t_{\rm FWHM}} \int\! \mathrm{d}m_r \dfrac{\mathrm{d}N_{\rm event}}{\mathrm{d} \log(t_{\rm FWHM})} \dfrac{\mathrm{d}N_s}{\mathrm{d}m_r} \epsilon(t_{\rm FWHM}, m_r) \dfrac{f_{\rm PBH}(M)}{M}~.
\end{equation}
\end{widetext}
Here, $(\Omega_{\rm PBH}/\Omega_{\rm DM}) = f_{\rm PBH,tot}$ is the 
{mass} fraction of DM in the form of PBHs, $\mathrm{d}N_{\rm event}/\mathrm{d} \log(t_{\rm FWHM})$ is the expected differential number of PBH microlensing events per logarithmic interval of the fullwidth-at-half-maximum (FWHM) microlensing timescale $t_{\rm FWHM}$ for a single star in M31, 
$\mathrm{d}N_s/\mathrm{d}m_r$ is the luminosity function of source stars in the photometric $r$-band magnitude range $[m_r, r + \mathrm{d}m_r]$, and  $\epsilon(t_{\rm FWHM}, m_r)$ is the detection efficiency (which quantifies the probability that a microlensing event for a star
with magnitude $m_r$ and light curve timescale $t_{\rm FWHM}$ is detected by HSC event selection procedures). We normalize the PBH mass function $f_{\rm PBH}(M)$ by enforcing the condition that PBHs constitute the entirety of DM, $\int_0^{\infty}dM f_{\rm PBH}(M)/M = f_{\rm PBH, tot}=1$.

The PBH mass function described by Eq.~\eqref{eq:mass_function}, with normalization and mass peak (at $\sim 10^{-14}M_\odot$) consistent with both PBH~DM and NANOGrav observations, is compatible at 2-$\sigma$ CL with the detection of a single candidate event reported by HSC after 7 hours of observations (at $M \sim 10^{-9} M_{\odot}$), and zero events at any other mass. The 95\% CL allowed range for the normalization of the mass function  $f_{\rm PBH}(M) \propto M^{-1/2}$ which satisfies these conditions is shown in purple in the left panel of Fig.~\ref{fig:PBHfull}. Notice that as the PBH spectrum is not monochromatic, the allowed range does not reach the differential HSC exclusion region for a monochromatic mass function.

Based on Poisson statistics, detecting a single event after the 7 hours of observations already carried out  by HSC is compatible with the PBH~DM hypothesis and a spectrum scaling as $M^{-1/2}$ at $\sim 16\%$ CL. New detections are expected with only 2.4 hours of observations. 
The scenario of $f_{\rm PBH,tot}=1$ and spectrum $f_{\rm PBH}(M_{\rm PBH})$ given by Eq.~\eqref{eq:mass_function} can be excluded  at a 2-$\sigma$ level (95\% CL) with additional 4 hours of observation, when combined with the existing 7 hours of observation and assuming null event detection in future observations. The red shaded region in right panel of Fig.~\ref{fig:PBHfull} is the exclusion region after 11 hours of observation in total.

Since solar-mass PBHs are associated with 2nd order curvature perturbations responsible for the NANOGrav signal, a direct probe of this mass-range would be desirable. Solar-mass PBHs can be probed by longer duration microlensing observations, such as MACHO and OGLE. However, in this mass range, stars can also contribute to microlensing events and it’s more difficult to distinguish contributions of PBHs. Caustic crossings \cite{Oguri:2017ock} as well as LIGO merger GWs (e.g. \cite{Ali-Haimoud:2017rtz}), which could probe the solar-mass PBH region, are also not expected to be sufficiently sensitive at the abundance that we consider.

In conclusion, we have highlighted that SGWB signals from PBHs with an extended mass function and comprising DM can be well probed by optical telescopes, independently of other GW-specific experiments. In particular, SGWB from PBH DM formation due to a broad perturbation spectrum and consistent with the signal detected by the NANOGrav collaboration is already being tested by HSC. Near future observations by HSC and other optical telescopes will provide a definitive independent test of this possibility.

The work of A.K., V.T. and E.V. was supported by the U.S. Department of Energy (DOE) Grant No. DE-SC0009937.  A.K. and V.T. were also supported by the World Premier International Research Center Initiative (WPI), MEXT, Japan. The work of S.S. was supported by International Graduate Program for Excellence in Earth-Space Science (IGPEES), World- leading Innovative Graduate Study (WINGS) Program, the University of Tokyo. This research was also supported in part by the National Science Foundation under Grant No. NSF PHY-1748958. This work was supported in part by JSPS KAKENHI Grant Numbers JP15H03654, JP15H05887, JP15H05893, JP15H05896, JP15K21733, JP19H00677, JP19H01895 and JP20H04727. 

 
\bibliography{bibliography}

 \newcommand{\noop}[1]{}
\begin{thebibliography}{104}%
\makeatletter
\providecommand \@ifxundefined [1]{%
 \@ifx{#1\undefined}
}%
\providecommand \@ifnum [1]{%
 \ifnum #1\expandafter \@firstoftwo
 \else \expandafter \@secondoftwo
 \fi
}%
\providecommand \@ifx [1]{%
 \ifx #1\expandafter \@firstoftwo
 \else \expandafter \@secondoftwo
 \fi
}%
\providecommand \natexlab [1]{#1}%
\providecommand \enquote  [1]{``#1''}%
\providecommand \bibnamefont  [1]{#1}%
\providecommand \bibfnamefont [1]{#1}%
\providecommand \citenamefont [1]{#1}%
\providecommand \href@noop [0]{\@secondoftwo}%
\providecommand \href [0]{\begingroup \@sanitize@url \@href}%
\providecommand \@href[1]{\@@startlink{#1}\@@href}%
\providecommand \@@href[1]{\endgroup#1\@@endlink}%
\providecommand \@sanitize@url [0]{\catcode `\\12\catcode `\$12\catcode
  `\&12\catcode `\#12\catcode `\^12\catcode `\_12\catcode `\%12\relax}%
\providecommand \@@startlink[1]{}%
\providecommand \@@endlink[0]{}%
\providecommand \url  [0]{\begingroup\@sanitize@url \@url }%
\providecommand \@url [1]{\endgroup\@href {#1}{\urlprefix }}%
\providecommand \urlprefix  [0]{URL }%
\providecommand \Eprint [0]{\href }%
\providecommand \doibase [0]{http://dx.doi.org/}%
\providecommand \selectlanguage [0]{\@gobble}%
\providecommand \bibinfo  [0]{\@secondoftwo}%
\providecommand \bibfield  [0]{\@secondoftwo}%
\providecommand \translation [1]{[#1]}%
\providecommand \BibitemOpen [0]{}%
\providecommand \bibitemStop [0]{}%
\providecommand \bibitemNoStop [0]{.\EOS\space}%
\providecommand \EOS [0]{\spacefactor3000\relax}%
\providecommand \BibitemShut  [1]{\csname bibitem#1\endcsname}%
\let\auto@bib@innerbib\@empty
\bibitem [{\citenamefont {{Zel'dovich}}\ and\ \citenamefont
  {{Novikov}}(1967)}]{Zeldovich:1967}%
  \BibitemOpen
  \bibfield  {author} {\bibinfo {author} {\bibfnamefont {Y.~B.}\ \bibnamefont
  {{Zel'dovich}}}\ and\ \bibinfo {author} {\bibfnamefont {I.~D.}\ \bibnamefont
  {{Novikov}}},\ }\href@noop {} {\bibfield  {journal} {\bibinfo  {journal}
  {Sov. Astron.}\ }\textbf {\bibinfo {volume} {10}},\ \bibinfo {pages} {602}
  (\bibinfo {year} {1967})}\BibitemShut {NoStop}%
\bibitem [{\citenamefont {Hawking}(1971)}]{Hawking:1971ei}%
  \BibitemOpen
  \bibfield  {author} {\bibinfo {author} {\bibfnamefont {S.}~\bibnamefont
  {Hawking}},\ }\href@noop {} {\bibfield  {journal} {\bibinfo  {journal} {Mon.
  Not. Roy. Astron. Soc.}\ }\textbf {\bibinfo {volume} {152}},\ \bibinfo
  {pages} {75} (\bibinfo {year} {1971})}\BibitemShut {NoStop}%
\bibitem [{\citenamefont {Carr}\ and\ \citenamefont
  {Hawking}(1974)}]{Carr:1974nx}%
  \BibitemOpen
  \bibfield  {author} {\bibinfo {author} {\bibfnamefont {B.~J.}\ \bibnamefont
  {Carr}}\ and\ \bibinfo {author} {\bibfnamefont {S.~W.}\ \bibnamefont
  {Hawking}},\ }\href@noop {} {\bibfield  {journal} {\bibinfo  {journal} {Mon.
  Not. Roy. Astron. Soc.}\ }\textbf {\bibinfo {volume} {168}},\ \bibinfo
  {pages} {399} (\bibinfo {year} {1974})}\BibitemShut {NoStop}%
\bibitem [{\citenamefont {Garcia-Bellido}\ \emph {et~al.}(1996)\citenamefont
  {Garcia-Bellido}, \citenamefont {Linde},\ and\ \citenamefont
  {Wands}}]{GarciaBellido:1996qt}%
  \BibitemOpen
  \bibfield  {author} {\bibinfo {author} {\bibfnamefont {J.}~\bibnamefont
  {Garcia-Bellido}}, \bibinfo {author} {\bibfnamefont {A.~D.}\ \bibnamefont
  {Linde}}, \ and\ \bibinfo {author} {\bibfnamefont {D.}~\bibnamefont
  {Wands}},\ }\href {\doibase 10.1103/PhysRevD.54.6040} {\bibfield  {journal}
  {\bibinfo  {journal} {Phys. Rev.}\ }\textbf {\bibinfo {volume} {D54}},\
  \bibinfo {pages} {6040} (\bibinfo {year} {1996})},\ \Eprint
  {http://arxiv.org/abs/astro-ph/9605094} {arXiv:astro-ph/9605094 [astro-ph]}
  \BibitemShut {NoStop}%
\bibitem [{\citenamefont {Green}\ and\ \citenamefont
  {Malik}(2001)}]{Green:2000he}%
  \BibitemOpen
  \bibfield  {author} {\bibinfo {author} {\bibfnamefont {A.~M.}\ \bibnamefont
  {Green}}\ and\ \bibinfo {author} {\bibfnamefont {K.~A.}\ \bibnamefont
  {Malik}},\ }\href {\doibase 10.1103/PhysRevD.64.021301} {\bibfield  {journal}
  {\bibinfo  {journal} {Phys. Rev. D}\ }\textbf {\bibinfo {volume} {64}},\
  \bibinfo {pages} {021301} (\bibinfo {year} {2001})},\ \Eprint
  {http://arxiv.org/abs/hep-ph/0008113} {arXiv:hep-ph/0008113} \BibitemShut
  {NoStop}%
\bibitem [{\citenamefont {Khlopov}(2010)}]{Khlopov:2008qy}%
  \BibitemOpen
  \bibfield  {author} {\bibinfo {author} {\bibfnamefont {M.~{\relax Yu}.}\
  \bibnamefont {Khlopov}},\ }\href {\doibase 10.1088/1674-4527/10/6/001}
  {\bibfield  {journal} {\bibinfo  {journal} {Res. Astron. Astrophys.}\
  }\textbf {\bibinfo {volume} {10}},\ \bibinfo {pages} {495} (\bibinfo {year}
  {2010})},\ \Eprint {http://arxiv.org/abs/0801.0116} {arXiv:0801.0116
  [astro-ph]} \BibitemShut {NoStop}%
\bibitem [{\citenamefont {Frampton}\ \emph {et~al.}(2010)\citenamefont
  {Frampton}, \citenamefont {Kawasaki}, \citenamefont {Takahashi},\ and\
  \citenamefont {Yanagida}}]{Frampton:2010sw}%
  \BibitemOpen
  \bibfield  {author} {\bibinfo {author} {\bibfnamefont {P.~H.}\ \bibnamefont
  {Frampton}}, \bibinfo {author} {\bibfnamefont {M.}~\bibnamefont {Kawasaki}},
  \bibinfo {author} {\bibfnamefont {F.}~\bibnamefont {Takahashi}}, \ and\
  \bibinfo {author} {\bibfnamefont {T.~T.}\ \bibnamefont {Yanagida}},\ }\href
  {\doibase 10.1088/1475-7516/2010/04/023} {\bibfield  {journal} {\bibinfo
  {journal} {JCAP}\ }\textbf {\bibinfo {volume} {1004}},\ \bibinfo {pages}
  {023} (\bibinfo {year} {2010})},\ \Eprint {http://arxiv.org/abs/1001.2308}
  {arXiv:1001.2308 [hep-ph]} \BibitemShut {NoStop}%
\bibitem [{\citenamefont {Kawasaki}\ \emph {et~al.}(2016)\citenamefont
  {Kawasaki}, \citenamefont {Kusenko}, \citenamefont {Tada},\ and\
  \citenamefont {Yanagida}}]{Kawasaki:2016pql}%
  \BibitemOpen
  \bibfield  {author} {\bibinfo {author} {\bibfnamefont {M.}~\bibnamefont
  {Kawasaki}}, \bibinfo {author} {\bibfnamefont {A.}~\bibnamefont {Kusenko}},
  \bibinfo {author} {\bibfnamefont {Y.}~\bibnamefont {Tada}}, \ and\ \bibinfo
  {author} {\bibfnamefont {T.~T.}\ \bibnamefont {Yanagida}},\ }\href {\doibase
  10.1103/PhysRevD.94.083523} {\bibfield  {journal} {\bibinfo  {journal} {Phys.
  Rev.}\ }\textbf {\bibinfo {volume} {D94}},\ \bibinfo {pages} {083523}
  (\bibinfo {year} {2016})},\ \Eprint {http://arxiv.org/abs/1606.07631}
  {arXiv:1606.07631 [astro-ph.CO]} \BibitemShut {NoStop}%
\bibitem [{\citenamefont {Cotner}\ and\ \citenamefont
  {Kusenko}(2017{\natexlab{a}})}]{Cotner:2016cvr}%
  \BibitemOpen
  \bibfield  {author} {\bibinfo {author} {\bibfnamefont {E.}~\bibnamefont
  {Cotner}}\ and\ \bibinfo {author} {\bibfnamefont {A.}~\bibnamefont
  {Kusenko}},\ }\href {\doibase 10.1103/PhysRevLett.119.031103} {\bibfield
  {journal} {\bibinfo  {journal} {Phys. Rev. Lett.}\ }\textbf {\bibinfo
  {volume} {119}},\ \bibinfo {pages} {031103} (\bibinfo {year}
  {2017}{\natexlab{a}})},\ \Eprint {http://arxiv.org/abs/1612.02529}
  {arXiv:1612.02529 [astro-ph.CO]} \BibitemShut {NoStop}%
\bibitem [{\citenamefont {Cotner}\ and\ \citenamefont
  {Kusenko}(2017{\natexlab{b}})}]{Cotner:2017tir}%
  \BibitemOpen
  \bibfield  {author} {\bibinfo {author} {\bibfnamefont {E.}~\bibnamefont
  {Cotner}}\ and\ \bibinfo {author} {\bibfnamefont {A.}~\bibnamefont
  {Kusenko}},\ }\href {\doibase 10.1103/PhysRevD.96.103002} {\bibfield
  {journal} {\bibinfo  {journal} {Phys. Rev.}\ }\textbf {\bibinfo {volume}
  {D96}},\ \bibinfo {pages} {103002} (\bibinfo {year} {2017}{\natexlab{b}})},\
  \Eprint {http://arxiv.org/abs/1706.09003} {arXiv:1706.09003 [astro-ph.CO]}
  \BibitemShut {NoStop}%
\bibitem [{\citenamefont {Green}(2016)}]{Green:2016xgy}%
  \BibitemOpen
  \bibfield  {author} {\bibinfo {author} {\bibfnamefont {A.~M.}\ \bibnamefont
  {Green}},\ }\href {\doibase 10.1103/PhysRevD.94.063530} {\bibfield  {journal}
  {\bibinfo  {journal} {Phys. Rev. D}\ }\textbf {\bibinfo {volume} {94}},\
  \bibinfo {pages} {063530} (\bibinfo {year} {2016})},\ \Eprint
  {http://arxiv.org/abs/1609.01143} {arXiv:1609.01143 [astro-ph.CO]}
  \BibitemShut {NoStop}%
\bibitem [{\citenamefont {Carr}\ \emph {et~al.}(2016)\citenamefont {Carr},
  \citenamefont {Kuhnel},\ and\ \citenamefont {Sandstad}}]{Carr:2016drx}%
  \BibitemOpen
  \bibfield  {author} {\bibinfo {author} {\bibfnamefont {B.}~\bibnamefont
  {Carr}}, \bibinfo {author} {\bibfnamefont {F.}~\bibnamefont {Kuhnel}}, \ and\
  \bibinfo {author} {\bibfnamefont {M.}~\bibnamefont {Sandstad}},\ }\href
  {\doibase 10.1103/PhysRevD.94.083504} {\bibfield  {journal} {\bibinfo
  {journal} {Phys. Rev.}\ }\textbf {\bibinfo {volume} {D94}},\ \bibinfo {pages}
  {083504} (\bibinfo {year} {2016})},\ \Eprint
  {http://arxiv.org/abs/1607.06077} {arXiv:1607.06077 [astro-ph.CO]}
  \BibitemShut {NoStop}%
\bibitem [{\citenamefont {Inomata}\ \emph {et~al.}(2016)\citenamefont
  {Inomata}, \citenamefont {Kawasaki}, \citenamefont {Mukaida}, \citenamefont
  {Tada},\ and\ \citenamefont {Yanagida}}]{Inomata:2016rbd}%
  \BibitemOpen
  \bibfield  {author} {\bibinfo {author} {\bibfnamefont {K.}~\bibnamefont
  {Inomata}}, \bibinfo {author} {\bibfnamefont {M.}~\bibnamefont {Kawasaki}},
  \bibinfo {author} {\bibfnamefont {K.}~\bibnamefont {Mukaida}}, \bibinfo
  {author} {\bibfnamefont {Y.}~\bibnamefont {Tada}}, \ and\ \bibinfo {author}
  {\bibfnamefont {T.~T.}\ \bibnamefont {Yanagida}},\ }\href@noop {} {\
  (\bibinfo {year} {2016})},\ \Eprint {http://arxiv.org/abs/1611.06130}
  {arXiv:1611.06130 [astro-ph.CO]} \BibitemShut {NoStop}%
\bibitem [{\citenamefont {Pi}\ \emph {et~al.}(2018)\citenamefont {Pi},
  \citenamefont {Zhang}, \citenamefont {Huang},\ and\ \citenamefont
  {Sasaki}}]{Pi:2017gih}%
  \BibitemOpen
  \bibfield  {author} {\bibinfo {author} {\bibfnamefont {S.}~\bibnamefont
  {Pi}}, \bibinfo {author} {\bibfnamefont {Y.-l.}\ \bibnamefont {Zhang}},
  \bibinfo {author} {\bibfnamefont {Q.-G.}\ \bibnamefont {Huang}}, \ and\
  \bibinfo {author} {\bibfnamefont {M.}~\bibnamefont {Sasaki}},\ }\href
  {\doibase 10.1088/1475-7516/2018/05/042} {\bibfield  {journal} {\bibinfo
  {journal} {JCAP}\ }\textbf {\bibinfo {volume} {1805}},\ \bibinfo {pages}
  {042} (\bibinfo {year} {2018})},\ \Eprint {http://arxiv.org/abs/1712.09896}
  {arXiv:1712.09896 [astro-ph.CO]} \BibitemShut {NoStop}%
\bibitem [{\citenamefont {Inomata}\ \emph
  {et~al.}(2017{\natexlab{a}})\citenamefont {Inomata}, \citenamefont
  {Kawasaki}, \citenamefont {Mukaida}, \citenamefont {Tada},\ and\
  \citenamefont {Yanagida}}]{Inomata:2017okj}%
  \BibitemOpen
  \bibfield  {author} {\bibinfo {author} {\bibfnamefont {K.}~\bibnamefont
  {Inomata}}, \bibinfo {author} {\bibfnamefont {M.}~\bibnamefont {Kawasaki}},
  \bibinfo {author} {\bibfnamefont {K.}~\bibnamefont {Mukaida}}, \bibinfo
  {author} {\bibfnamefont {Y.}~\bibnamefont {Tada}}, \ and\ \bibinfo {author}
  {\bibfnamefont {T.~T.}\ \bibnamefont {Yanagida}},\ }\href@noop {} {\
  (\bibinfo {year} {2017}{\natexlab{a}})},\ \Eprint
  {http://arxiv.org/abs/1701.02544} {arXiv:1701.02544 [astro-ph.CO]}
  \BibitemShut {NoStop}%
\bibitem [{\citenamefont {Garcia-Bellido}\ \emph {et~al.}(2017)\citenamefont
  {Garcia-Bellido}, \citenamefont {Peloso},\ and\ \citenamefont
  {Unal}}]{Garcia-Bellido:2017aan}%
  \BibitemOpen
  \bibfield  {author} {\bibinfo {author} {\bibfnamefont {J.}~\bibnamefont
  {Garcia-Bellido}}, \bibinfo {author} {\bibfnamefont {M.}~\bibnamefont
  {Peloso}}, \ and\ \bibinfo {author} {\bibfnamefont {C.}~\bibnamefont
  {Unal}},\ }\href {\doibase 10.1088/1475-7516/2017/09/013} {\bibfield
  {journal} {\bibinfo  {journal} {JCAP}\ }\textbf {\bibinfo {volume} {1709}},\
  \bibinfo {pages} {013} (\bibinfo {year} {2017})},\ \Eprint
  {http://arxiv.org/abs/1707.02441} {arXiv:1707.02441 [astro-ph.CO]}
  \BibitemShut {NoStop}%
\bibitem [{\citenamefont {Inoue}\ and\ \citenamefont
  {Kusenko}(2017)}]{Inoue:2017csr}%
  \BibitemOpen
  \bibfield  {author} {\bibinfo {author} {\bibfnamefont {Y.}~\bibnamefont
  {Inoue}}\ and\ \bibinfo {author} {\bibfnamefont {A.}~\bibnamefont
  {Kusenko}},\ }\href {\doibase 10.1088/1475-7516/2017/10/034} {\bibfield
  {journal} {\bibinfo  {journal} {JCAP}\ }\textbf {\bibinfo {volume} {1710}},\
  \bibinfo {pages} {034} (\bibinfo {year} {2017})},\ \Eprint
  {http://arxiv.org/abs/1705.00791} {arXiv:1705.00791 [astro-ph.CO]}
  \BibitemShut {NoStop}%
\bibitem [{\citenamefont {Georg}\ and\ \citenamefont
  {Watson}(2017)}]{Georg:2017mqk}%
  \BibitemOpen
  \bibfield  {author} {\bibinfo {author} {\bibfnamefont {J.}~\bibnamefont
  {Georg}}\ and\ \bibinfo {author} {\bibfnamefont {S.}~\bibnamefont {Watson}},\
  }\href@noop {} {\  (\bibinfo {year} {2017})},\ \Eprint
  {http://arxiv.org/abs/1703.04825} {arXiv:1703.04825 [astro-ph.CO]}
  \BibitemShut {NoStop}%
\bibitem [{\citenamefont {Inomata}\ \emph
  {et~al.}(2017{\natexlab{b}})\citenamefont {Inomata}, \citenamefont
  {Kawasaki}, \citenamefont {Mukaida},\ and\ \citenamefont
  {Yanagida}}]{Inomata:2017bwi}%
  \BibitemOpen
  \bibfield  {author} {\bibinfo {author} {\bibfnamefont {K.}~\bibnamefont
  {Inomata}}, \bibinfo {author} {\bibfnamefont {M.}~\bibnamefont {Kawasaki}},
  \bibinfo {author} {\bibfnamefont {K.}~\bibnamefont {Mukaida}}, \ and\
  \bibinfo {author} {\bibfnamefont {T.~T.}\ \bibnamefont {Yanagida}},\
  }\href@noop {} {\  (\bibinfo {year} {2017}{\natexlab{b}})},\ \Eprint
  {http://arxiv.org/abs/1711.06129} {arXiv:1711.06129 [astro-ph.CO]}
  \BibitemShut {NoStop}%
\bibitem [{\citenamefont {Kocsis}\ \emph {et~al.}(2017)\citenamefont {Kocsis},
  \citenamefont {Suyama}, \citenamefont {Tanaka},\ and\ \citenamefont
  {Yokoyama}}]{Kocsis:2017yty}%
  \BibitemOpen
  \bibfield  {author} {\bibinfo {author} {\bibfnamefont {B.}~\bibnamefont
  {Kocsis}}, \bibinfo {author} {\bibfnamefont {T.}~\bibnamefont {Suyama}},
  \bibinfo {author} {\bibfnamefont {T.}~\bibnamefont {Tanaka}}, \ and\ \bibinfo
  {author} {\bibfnamefont {S.}~\bibnamefont {Yokoyama}},\ }\href@noop {} {\
  (\bibinfo {year} {2017})},\ \Eprint {http://arxiv.org/abs/1709.09007}
  {arXiv:1709.09007 [astro-ph.CO]} \BibitemShut {NoStop}%
\bibitem [{\citenamefont {Ando}\ \emph {et~al.}(2017)\citenamefont {Ando},
  \citenamefont {Inomata}, \citenamefont {Kawasaki}, \citenamefont {Mukaida},\
  and\ \citenamefont {Yanagida}}]{Ando:2017veq}%
  \BibitemOpen
  \bibfield  {author} {\bibinfo {author} {\bibfnamefont {K.}~\bibnamefont
  {Ando}}, \bibinfo {author} {\bibfnamefont {K.}~\bibnamefont {Inomata}},
  \bibinfo {author} {\bibfnamefont {M.}~\bibnamefont {Kawasaki}}, \bibinfo
  {author} {\bibfnamefont {K.}~\bibnamefont {Mukaida}}, \ and\ \bibinfo
  {author} {\bibfnamefont {T.~T.}\ \bibnamefont {Yanagida}},\ }\href@noop {} {\
   (\bibinfo {year} {2017})},\ \Eprint {http://arxiv.org/abs/1711.08956}
  {arXiv:1711.08956 [astro-ph.CO]} \BibitemShut {NoStop}%
\bibitem [{\citenamefont {Germani}\ and\ \citenamefont
  {Prokopec}(2017)}]{Germani:2017bcs}%
  \BibitemOpen
  \bibfield  {author} {\bibinfo {author} {\bibfnamefont {C.}~\bibnamefont
  {Germani}}\ and\ \bibinfo {author} {\bibfnamefont {T.}~\bibnamefont
  {Prokopec}},\ }\href {\doibase 10.1016/j.dark.2017.09.001} {\bibfield
  {journal} {\bibinfo  {journal} {Phys. Dark Univ.}\ }\textbf {\bibinfo
  {volume} {18}},\ \bibinfo {pages} {6} (\bibinfo {year} {2017})},\ \Eprint
  {http://arxiv.org/abs/1706.04226} {arXiv:1706.04226 [astro-ph.CO]}
  \BibitemShut {NoStop}%
\bibitem [{\citenamefont {Nishikawa}\ \emph {et~al.}(2019)\citenamefont
  {Nishikawa}, \citenamefont {Kovetz}, \citenamefont {Kamionkowski},\ and\
  \citenamefont {Silk}}]{Nishikawa:2017chy}%
  \BibitemOpen
  \bibfield  {author} {\bibinfo {author} {\bibfnamefont {H.}~\bibnamefont
  {Nishikawa}}, \bibinfo {author} {\bibfnamefont {E.~D.}\ \bibnamefont
  {Kovetz}}, \bibinfo {author} {\bibfnamefont {M.}~\bibnamefont
  {Kamionkowski}}, \ and\ \bibinfo {author} {\bibfnamefont {J.}~\bibnamefont
  {Silk}},\ }\href {\doibase 10.1103/PhysRevD.99.043533} {\bibfield  {journal}
  {\bibinfo  {journal} {Phys. Rev.}\ }\textbf {\bibinfo {volume} {D99}},\
  \bibinfo {pages} {043533} (\bibinfo {year} {2019})},\ \Eprint
  {http://arxiv.org/abs/1708.08449} {arXiv:1708.08449 [astro-ph.CO]}
  \BibitemShut {NoStop}%
\bibitem [{\citenamefont {Germani}\ and\ \citenamefont
  {Musco}(2019)}]{Germani:2018jgr}%
  \BibitemOpen
  \bibfield  {author} {\bibinfo {author} {\bibfnamefont {C.}~\bibnamefont
  {Germani}}\ and\ \bibinfo {author} {\bibfnamefont {I.}~\bibnamefont
  {Musco}},\ }\href {\doibase 10.1103/PhysRevLett.122.141302} {\bibfield
  {journal} {\bibinfo  {journal} {Phys. Rev. Lett.}\ }\textbf {\bibinfo
  {volume} {122}},\ \bibinfo {pages} {141302} (\bibinfo {year} {2019})},\
  \Eprint {http://arxiv.org/abs/1805.04087} {arXiv:1805.04087 [astro-ph.CO]}
  \BibitemShut {NoStop}%
\bibitem [{\citenamefont {Cotner}\ \emph {et~al.}(2018)\citenamefont {Cotner},
  \citenamefont {Kusenko},\ and\ \citenamefont {Takhistov}}]{Cotner:2018vug}%
  \BibitemOpen
  \bibfield  {author} {\bibinfo {author} {\bibfnamefont {E.}~\bibnamefont
  {Cotner}}, \bibinfo {author} {\bibfnamefont {A.}~\bibnamefont {Kusenko}}, \
  and\ \bibinfo {author} {\bibfnamefont {V.}~\bibnamefont {Takhistov}},\ }\href
  {\doibase 10.1103/PhysRevD.98.083513} {\bibfield  {journal} {\bibinfo
  {journal} {Phys. Rev.}\ }\textbf {\bibinfo {volume} {D98}},\ \bibinfo {pages}
  {083513} (\bibinfo {year} {2018})},\ \Eprint
  {http://arxiv.org/abs/1801.03321} {arXiv:1801.03321 [astro-ph.CO]}
  \BibitemShut {NoStop}%
\bibitem [{\citenamefont {Sasaki}\ \emph {et~al.}(2018)\citenamefont {Sasaki},
  \citenamefont {Suyama}, \citenamefont {Tanaka},\ and\ \citenamefont
  {Yokoyama}}]{Sasaki:2018dmp}%
  \BibitemOpen
  \bibfield  {author} {\bibinfo {author} {\bibfnamefont {M.}~\bibnamefont
  {Sasaki}}, \bibinfo {author} {\bibfnamefont {T.}~\bibnamefont {Suyama}},
  \bibinfo {author} {\bibfnamefont {T.}~\bibnamefont {Tanaka}}, \ and\ \bibinfo
  {author} {\bibfnamefont {S.}~\bibnamefont {Yokoyama}},\ }\href {\doibase
  10.1088/1361-6382/aaa7b4} {\bibfield  {journal} {\bibinfo  {journal} {Class.
  Quant. Grav.}\ }\textbf {\bibinfo {volume} {35}},\ \bibinfo {pages} {063001}
  (\bibinfo {year} {2018})},\ \Eprint {http://arxiv.org/abs/1801.05235}
  {arXiv:1801.05235 [astro-ph.CO]} \BibitemShut {NoStop}%
\bibitem [{\citenamefont {Carr}\ and\ \citenamefont
  {Silk}(2018)}]{Carr:2018rid}%
  \BibitemOpen
  \bibfield  {author} {\bibinfo {author} {\bibfnamefont {B.}~\bibnamefont
  {Carr}}\ and\ \bibinfo {author} {\bibfnamefont {J.}~\bibnamefont {Silk}},\
  }\href {\doibase 10.1093/mnras/sty1204} {\bibfield  {journal} {\bibinfo
  {journal} {Mon. Not. Roy. Astron. Soc.}\ }\textbf {\bibinfo {volume} {478}},\
  \bibinfo {pages} {3756} (\bibinfo {year} {2018})},\ \Eprint
  {http://arxiv.org/abs/1801.00672} {arXiv:1801.00672 [astro-ph.CO]}
  \BibitemShut {NoStop}%
\bibitem [{\citenamefont {Cotner}\ \emph {et~al.}(2019)\citenamefont {Cotner},
  \citenamefont {Kusenko}, \citenamefont {Sasaki},\ and\ \citenamefont
  {Takhistov}}]{Cotner:2019ykd}%
  \BibitemOpen
  \bibfield  {author} {\bibinfo {author} {\bibfnamefont {E.}~\bibnamefont
  {Cotner}}, \bibinfo {author} {\bibfnamefont {A.}~\bibnamefont {Kusenko}},
  \bibinfo {author} {\bibfnamefont {M.}~\bibnamefont {Sasaki}}, \ and\ \bibinfo
  {author} {\bibfnamefont {V.}~\bibnamefont {Takhistov}},\ }\href {\doibase
  10.1088/1475-7516/2019/10/077} {\bibfield  {journal} {\bibinfo  {journal}
  {JCAP}\ }\textbf {\bibinfo {volume} {1910}},\ \bibinfo {pages} {077}
  (\bibinfo {year} {2019})},\ \Eprint {http://arxiv.org/abs/1907.10613}
  {arXiv:1907.10613 [astro-ph.CO]} \BibitemShut {NoStop}%
\bibitem [{\citenamefont {Germani}\ and\ \citenamefont
  {Sheth}(2020)}]{Germani:2019zez}%
  \BibitemOpen
  \bibfield  {author} {\bibinfo {author} {\bibfnamefont {C.}~\bibnamefont
  {Germani}}\ and\ \bibinfo {author} {\bibfnamefont {R.~K.}\ \bibnamefont
  {Sheth}},\ }\href {\doibase 10.1103/PhysRevD.101.063520} {\bibfield
  {journal} {\bibinfo  {journal} {Phys. Rev.}\ }\textbf {\bibinfo {volume}
  {D101}},\ \bibinfo {pages} {063520} (\bibinfo {year} {2020})},\ \Eprint
  {http://arxiv.org/abs/1912.07072} {arXiv:1912.07072 [astro-ph.CO]}
  \BibitemShut {NoStop}%
\bibitem [{\citenamefont {Inman}\ and\ \citenamefont
  {Ali-Haïmoud}(2019)}]{Inman:2019wvr}%
  \BibitemOpen
  \bibfield  {author} {\bibinfo {author} {\bibfnamefont {D.}~\bibnamefont
  {Inman}}\ and\ \bibinfo {author} {\bibfnamefont {Y.}~\bibnamefont
  {Ali-Haïmoud}},\ }\href {\doibase 10.1103/PhysRevD.100.083528} {\bibfield
  {journal} {\bibinfo  {journal} {Phys. Rev.}\ }\textbf {\bibinfo {volume}
  {D100}},\ \bibinfo {pages} {083528} (\bibinfo {year} {2019})},\ \Eprint
  {http://arxiv.org/abs/1907.08129} {arXiv:1907.08129 [astro-ph.CO]}
  \BibitemShut {NoStop}%
\bibitem [{\citenamefont {Flores}\ and\ \citenamefont
  {Kusenko}(2020)}]{Flores:2020drq}%
  \BibitemOpen
  \bibfield  {author} {\bibinfo {author} {\bibfnamefont {M.~M.}\ \bibnamefont
  {Flores}}\ and\ \bibinfo {author} {\bibfnamefont {A.}~\bibnamefont
  {Kusenko}},\ }\href@noop {} {\  (\bibinfo {year} {2020})},\ \Eprint
  {http://arxiv.org/abs/2008.12456} {arXiv:2008.12456 [astro-ph.CO]}
  \BibitemShut {NoStop}%
\bibitem [{\citenamefont {Deng}\ and\ \citenamefont
  {Vilenkin}(2017)}]{Deng:2017uwc}%
  \BibitemOpen
  \bibfield  {author} {\bibinfo {author} {\bibfnamefont {H.}~\bibnamefont
  {Deng}}\ and\ \bibinfo {author} {\bibfnamefont {A.}~\bibnamefont
  {Vilenkin}},\ }\href {\doibase 10.1088/1475-7516/2017/12/044} {\bibfield
  {journal} {\bibinfo  {journal} {JCAP}\ }\textbf {\bibinfo {volume} {1712}},\
  \bibinfo {pages} {044} (\bibinfo {year} {2017})},\ \Eprint
  {http://arxiv.org/abs/1710.02865} {arXiv:1710.02865 [gr-qc]} \BibitemShut
  {NoStop}%
\bibitem [{\citenamefont {Green}\ and\ \citenamefont
  {Kavanagh}(2020)}]{Green:2020jor}%
  \BibitemOpen
  \bibfield  {author} {\bibinfo {author} {\bibfnamefont {A.~M.}\ \bibnamefont
  {Green}}\ and\ \bibinfo {author} {\bibfnamefont {B.~J.}\ \bibnamefont
  {Kavanagh}},\ }\href@noop {} {\  (\bibinfo {year} {2020})},\ \Eprint
  {http://arxiv.org/abs/2007.10722} {arXiv:2007.10722 [astro-ph.CO]}
  \BibitemShut {NoStop}%
\bibitem [{\citenamefont {Kusenko}\ \emph {et~al.}(2020)\citenamefont
  {Kusenko}, \citenamefont {Sasaki}, \citenamefont {Sugiyama}, \citenamefont
  {Takada}, \citenamefont {Takhistov},\ and\ \citenamefont
  {Vitagliano}}]{Kusenko:2020pcg}%
  \BibitemOpen
  \bibfield  {author} {\bibinfo {author} {\bibfnamefont {A.}~\bibnamefont
  {Kusenko}}, \bibinfo {author} {\bibfnamefont {M.}~\bibnamefont {Sasaki}},
  \bibinfo {author} {\bibfnamefont {S.}~\bibnamefont {Sugiyama}}, \bibinfo
  {author} {\bibfnamefont {M.}~\bibnamefont {Takada}}, \bibinfo {author}
  {\bibfnamefont {V.}~\bibnamefont {Takhistov}}, \ and\ \bibinfo {author}
  {\bibfnamefont {E.}~\bibnamefont {Vitagliano}},\ }\href@noop {} {\  (\bibinfo
  {year} {2020})},\ \Eprint {http://arxiv.org/abs/2001.09160} {arXiv:2001.09160
  [astro-ph.CO]} \BibitemShut {NoStop}%
\bibitem [{\citenamefont {de~Freitas~Pacheco}\ and\ \citenamefont
  {Silk}(2020)}]{deFreitasPacheco:2020wdg}%
  \BibitemOpen
  \bibfield  {author} {\bibinfo {author} {\bibfnamefont {J.~A.}\ \bibnamefont
  {de~Freitas~Pacheco}}\ and\ \bibinfo {author} {\bibfnamefont
  {J.}~\bibnamefont {Silk}},\ }\href {\doibase 10.1103/PhysRevD.101.083022}
  {\bibfield  {journal} {\bibinfo  {journal} {Phys. Rev.}\ }\textbf {\bibinfo
  {volume} {D101}},\ \bibinfo {pages} {083022} (\bibinfo {year} {2020})},\
  \Eprint {http://arxiv.org/abs/2003.12072} {arXiv:2003.12072 [astro-ph.CO]}
  \BibitemShut {NoStop}%
\bibitem [{\citenamefont {Nakamura}\ \emph {et~al.}(1997)\citenamefont
  {Nakamura}, \citenamefont {Sasaki}, \citenamefont {Tanaka},\ and\
  \citenamefont {Thorne}}]{Nakamura:1997sm}%
  \BibitemOpen
  \bibfield  {author} {\bibinfo {author} {\bibfnamefont {T.}~\bibnamefont
  {Nakamura}}, \bibinfo {author} {\bibfnamefont {M.}~\bibnamefont {Sasaki}},
  \bibinfo {author} {\bibfnamefont {T.}~\bibnamefont {Tanaka}}, \ and\ \bibinfo
  {author} {\bibfnamefont {K.~S.}\ \bibnamefont {Thorne}},\ }\href {\doibase
  10.1086/310886} {\bibfield  {journal} {\bibinfo  {journal} {Astrophys. J.}\
  }\textbf {\bibinfo {volume} {487}},\ \bibinfo {pages} {L139} (\bibinfo {year}
  {1997})},\ \Eprint {http://arxiv.org/abs/astro-ph/9708060}
  {arXiv:astro-ph/9708060 [astro-ph]} \BibitemShut {NoStop}%
\bibitem [{\citenamefont {Clesse}\ and\ \citenamefont
  {Garcia-Bellido}(2015)}]{Clesse:2015wea}%
  \BibitemOpen
  \bibfield  {author} {\bibinfo {author} {\bibfnamefont {S.}~\bibnamefont
  {Clesse}}\ and\ \bibinfo {author} {\bibfnamefont {J.}~\bibnamefont
  {Garcia-Bellido}},\ }\href {\doibase 10.1103/PhysRevD.92.023524} {\bibfield
  {journal} {\bibinfo  {journal} {Phys. Rev.}\ }\textbf {\bibinfo {volume}
  {D92}},\ \bibinfo {pages} {023524} (\bibinfo {year} {2015})},\ \Eprint
  {http://arxiv.org/abs/1501.07565} {arXiv:1501.07565 [astro-ph.CO]}
  \BibitemShut {NoStop}%
\bibitem [{\citenamefont {Bird}\ \emph {et~al.}(2016)\citenamefont {Bird} \emph
  {et~al.}}]{Bird:2016dcv}%
  \BibitemOpen
  \bibfield  {author} {\bibinfo {author} {\bibfnamefont {S.}~\bibnamefont
  {Bird}} \emph {et~al.},\ }\href {\doibase 10.1103/PhysRevLett.116.201301}
  {\bibfield  {journal} {\bibinfo  {journal} {Phys. Rev. Lett.}\ }\textbf
  {\bibinfo {volume} {116}},\ \bibinfo {pages} {201301} (\bibinfo {year}
  {2016})},\ \Eprint {http://arxiv.org/abs/1603.00464} {arXiv:1603.00464
  [astro-ph.CO]} \BibitemShut {NoStop}%
\bibitem [{\citenamefont {Raidal}\ \emph {et~al.}(2017)\citenamefont {Raidal},
  \citenamefont {Vaskonen},\ and\ \citenamefont {Veermae}}]{Raidal:2017mfl}%
  \BibitemOpen
  \bibfield  {author} {\bibinfo {author} {\bibfnamefont {M.}~\bibnamefont
  {Raidal}}, \bibinfo {author} {\bibfnamefont {V.}~\bibnamefont {Vaskonen}}, \
  and\ \bibinfo {author} {\bibfnamefont {H.}~\bibnamefont {Veermae}},\
  }\href@noop {} {\  (\bibinfo {year} {2017})},\ \Eprint
  {http://arxiv.org/abs/1707.01480} {arXiv:1707.01480 [astro-ph.CO]}
  \BibitemShut {NoStop}%
\bibitem [{\citenamefont {Eroshenko}(2016)}]{Eroshenko:2016hmn}%
  \BibitemOpen
  \bibfield  {author} {\bibinfo {author} {\bibfnamefont {{\relax Yu}.~N.}\
  \bibnamefont {Eroshenko}},\ }\href@noop {} {\  (\bibinfo {year} {2016})},\
  \Eprint {http://arxiv.org/abs/1604.04932} {arXiv:1604.04932 [astro-ph.CO]}
  \BibitemShut {NoStop}%
\bibitem [{\citenamefont {Sasaki}\ \emph {et~al.}(2016)\citenamefont {Sasaki},
  \citenamefont {Suyama}, \citenamefont {Tanaka},\ and\ \citenamefont
  {Yokoyama}}]{Sasaki:2016jop}%
  \BibitemOpen
  \bibfield  {author} {\bibinfo {author} {\bibfnamefont {M.}~\bibnamefont
  {Sasaki}}, \bibinfo {author} {\bibfnamefont {T.}~\bibnamefont {Suyama}},
  \bibinfo {author} {\bibfnamefont {T.}~\bibnamefont {Tanaka}}, \ and\ \bibinfo
  {author} {\bibfnamefont {S.}~\bibnamefont {Yokoyama}},\ }\href {\doibase
  10.1103/PhysRevLett.117.061101} {\bibfield  {journal} {\bibinfo  {journal}
  {Phys. Rev. Lett.}\ }\textbf {\bibinfo {volume} {117}},\ \bibinfo {pages}
  {061101} (\bibinfo {year} {2016})},\ \Eprint
  {http://arxiv.org/abs/1603.08338} {arXiv:1603.08338 [astro-ph.CO]}
  \BibitemShut {NoStop}%
\bibitem [{\citenamefont {Clesse}\ and\ \citenamefont
  {Garcia-Bellido}(2016)}]{Clesse:2016ajp}%
  \BibitemOpen
  \bibfield  {author} {\bibinfo {author} {\bibfnamefont {S.}~\bibnamefont
  {Clesse}}\ and\ \bibinfo {author} {\bibfnamefont {J.}~\bibnamefont
  {Garcia-Bellido}},\ }\href@noop {} {\  (\bibinfo {year} {2016})},\ \Eprint
  {http://arxiv.org/abs/1610.08479} {arXiv:1610.08479 [astro-ph.CO]}
  \BibitemShut {NoStop}%
\bibitem [{\citenamefont {Fuller}\ \emph {et~al.}(2017)\citenamefont {Fuller},
  \citenamefont {Kusenko},\ and\ \citenamefont {Takhistov}}]{Fuller:2017uyd}%
  \BibitemOpen
  \bibfield  {author} {\bibinfo {author} {\bibfnamefont {G.~M.}\ \bibnamefont
  {Fuller}}, \bibinfo {author} {\bibfnamefont {A.}~\bibnamefont {Kusenko}}, \
  and\ \bibinfo {author} {\bibfnamefont {V.}~\bibnamefont {Takhistov}},\ }\href
  {\doibase 10.1103/PhysRevLett.119.061101} {\bibfield  {journal} {\bibinfo
  {journal} {Phys. Rev. Lett.}\ }\textbf {\bibinfo {volume} {119}},\ \bibinfo
  {pages} {061101} (\bibinfo {year} {2017})},\ \Eprint
  {http://arxiv.org/abs/1704.01129} {arXiv:1704.01129 [astro-ph.HE]}
  \BibitemShut {NoStop}%
\bibitem [{\citenamefont {Abbott}\ \emph
  {et~al.}(2016{\natexlab{a}})\citenamefont {Abbott} \emph
  {et~al.}}]{Abbott:2016blz}%
  \BibitemOpen
  \bibfield  {author} {\bibinfo {author} {\bibfnamefont {B.~P.}\ \bibnamefont
  {Abbott}} \emph {et~al.} (\bibinfo {collaboration} {Virgo, LIGO
  Scientific}),\ }\href {\doibase 10.1103/PhysRevLett.116.061102} {\bibfield
  {journal} {\bibinfo  {journal} {Phys. Rev. Lett.}\ }\textbf {\bibinfo
  {volume} {116}},\ \bibinfo {pages} {061102} (\bibinfo {year}
  {2016}{\natexlab{a}})},\ \Eprint {http://arxiv.org/abs/1602.03837}
  {arXiv:1602.03837 [gr-qc]} \BibitemShut {NoStop}%
\bibitem [{\citenamefont {Abbott}\ \emph
  {et~al.}(2016{\natexlab{b}})\citenamefont {Abbott} \emph
  {et~al.}}]{Abbott:2016nmj}%
  \BibitemOpen
  \bibfield  {author} {\bibinfo {author} {\bibfnamefont {B.~P.}\ \bibnamefont
  {Abbott}} \emph {et~al.} (\bibinfo {collaboration} {Virgo, LIGO
  Scientific}),\ }\href {\doibase 10.1103/PhysRevLett.116.241103} {\bibfield
  {journal} {\bibinfo  {journal} {Phys. Rev. Lett.}\ }\textbf {\bibinfo
  {volume} {116}},\ \bibinfo {pages} {241103} (\bibinfo {year}
  {2016}{\natexlab{b}})},\ \Eprint {http://arxiv.org/abs/1606.04855}
  {arXiv:1606.04855 [gr-qc]} \BibitemShut {NoStop}%
\bibitem [{\citenamefont {Abbott}\ \emph {et~al.}(2017)\citenamefont {Abbott}
  \emph {et~al.}}]{Abbott:2017vtc}%
  \BibitemOpen
  \bibfield  {author} {\bibinfo {author} {\bibfnamefont {B.~P.}\ \bibnamefont
  {Abbott}} \emph {et~al.} (\bibinfo {collaboration} {VIRGO, LIGO
  Scientific}),\ }\href {\doibase 10.1103/PhysRevLett.118.221101} {\bibfield
  {journal} {\bibinfo  {journal} {Phys. Rev. Lett.}\ }\textbf {\bibinfo
  {volume} {118}},\ \bibinfo {pages} {221101} (\bibinfo {year} {2017})},\
  \Eprint {http://arxiv.org/abs/1706.01812} {arXiv:1706.01812 [gr-qc]}
  \BibitemShut {NoStop}%
\bibitem [{\citenamefont {Bean}\ and\ \citenamefont
  {Magueijo}(2002)}]{Bean:2002kx}%
  \BibitemOpen
  \bibfield  {author} {\bibinfo {author} {\bibfnamefont {R.}~\bibnamefont
  {Bean}}\ and\ \bibinfo {author} {\bibfnamefont {J.}~\bibnamefont
  {Magueijo}},\ }\href {\doibase 10.1103/PhysRevD.66.063505} {\bibfield
  {journal} {\bibinfo  {journal} {Phys. Rev.}\ }\textbf {\bibinfo {volume}
  {D66}},\ \bibinfo {pages} {063505} (\bibinfo {year} {2002})},\ \Eprint
  {http://arxiv.org/abs/astro-ph/0204486} {arXiv:astro-ph/0204486 [astro-ph]}
  \BibitemShut {NoStop}%
\bibitem [{\citenamefont {Kawasaki}\ \emph {et~al.}(2012)\citenamefont
  {Kawasaki}, \citenamefont {Kusenko},\ and\ \citenamefont
  {Yanagida}}]{Kawasaki:2012kn}%
  \BibitemOpen
  \bibfield  {author} {\bibinfo {author} {\bibfnamefont {M.}~\bibnamefont
  {Kawasaki}}, \bibinfo {author} {\bibfnamefont {A.}~\bibnamefont {Kusenko}}, \
  and\ \bibinfo {author} {\bibfnamefont {T.~T.}\ \bibnamefont {Yanagida}},\
  }\href {\doibase 10.1016/j.physletb.2012.03.056} {\bibfield  {journal}
  {\bibinfo  {journal} {Phys. Lett.}\ }\textbf {\bibinfo {volume} {B711}},\
  \bibinfo {pages} {1} (\bibinfo {year} {2012})},\ \Eprint
  {http://arxiv.org/abs/1202.3848} {arXiv:1202.3848 [astro-ph.CO]} \BibitemShut
  {NoStop}%
\bibitem [{\citenamefont {Takhistov}(2019)}]{Takhistov:2017nmt}%
  \BibitemOpen
  \bibfield  {author} {\bibinfo {author} {\bibfnamefont {V.}~\bibnamefont
  {Takhistov}},\ }\href {\doibase 10.1016/j.physletb.2018.12.043} {\bibfield
  {journal} {\bibinfo  {journal} {Phys. Lett.}\ }\textbf {\bibinfo {volume}
  {B789}},\ \bibinfo {pages} {538} (\bibinfo {year} {2019})},\ \Eprint
  {http://arxiv.org/abs/1710.09458} {arXiv:1710.09458 [astro-ph.HE]}
  \BibitemShut {NoStop}%
\bibitem [{\citenamefont {Takhistov}(2018)}]{Takhistov:2017bpt}%
  \BibitemOpen
  \bibfield  {author} {\bibinfo {author} {\bibfnamefont {V.}~\bibnamefont
  {Takhistov}},\ }\href {\doibase 10.1016/j.physletb.2018.05.026} {\bibfield
  {journal} {\bibinfo  {journal} {Phys. Lett.}\ }\textbf {\bibinfo {volume}
  {B782}},\ \bibinfo {pages} {77} (\bibinfo {year} {2018})},\ \Eprint
  {http://arxiv.org/abs/1707.05849} {arXiv:1707.05849 [astro-ph.CO]}
  \BibitemShut {NoStop}%
\bibitem [{\citenamefont {Katz}\ \emph {et~al.}(2018)\citenamefont {Katz},
  \citenamefont {Kopp}, \citenamefont {Sibiryakov},\ and\ \citenamefont
  {Xue}}]{Katz:2018zrn}%
  \BibitemOpen
  \bibfield  {author} {\bibinfo {author} {\bibfnamefont {A.}~\bibnamefont
  {Katz}}, \bibinfo {author} {\bibfnamefont {J.}~\bibnamefont {Kopp}}, \bibinfo
  {author} {\bibfnamefont {S.}~\bibnamefont {Sibiryakov}}, \ and\ \bibinfo
  {author} {\bibfnamefont {W.}~\bibnamefont {Xue}},\ }\href@noop {} {\bibfield
  {journal} {\bibinfo  {journal} {Submitted to: JCAP}\ } (\bibinfo {year}
  {2018})},\ \Eprint {http://arxiv.org/abs/1807.11495} {arXiv:1807.11495
  [astro-ph.CO]} \BibitemShut {NoStop}%
\bibitem [{\citenamefont {Smyth}\ \emph {et~al.}(2019)\citenamefont {Smyth},
  \citenamefont {Profumo}, \citenamefont {English}, \citenamefont {Jeltema},
  \citenamefont {McKinnon},\ and\ \citenamefont
  {Guhathakurta}}]{Smyth:2019whb}%
  \BibitemOpen
  \bibfield  {author} {\bibinfo {author} {\bibfnamefont {N.}~\bibnamefont
  {Smyth}}, \bibinfo {author} {\bibfnamefont {S.}~\bibnamefont {Profumo}},
  \bibinfo {author} {\bibfnamefont {S.}~\bibnamefont {English}}, \bibinfo
  {author} {\bibfnamefont {T.}~\bibnamefont {Jeltema}}, \bibinfo {author}
  {\bibfnamefont {K.}~\bibnamefont {McKinnon}}, \ and\ \bibinfo {author}
  {\bibfnamefont {P.}~\bibnamefont {Guhathakurta}},\ }\href@noop {} {\
  (\bibinfo {year} {2019})},\ \Eprint {http://arxiv.org/abs/1910.01285}
  {arXiv:1910.01285 [astro-ph.CO]} \BibitemShut {NoStop}%
\bibitem [{\citenamefont {Montero-Camacho}\ \emph {et~al.}(2019)\citenamefont
  {Montero-Camacho}, \citenamefont {Fang}, \citenamefont {Vasquez},
  \citenamefont {Silva},\ and\ \citenamefont
  {Hirata}}]{Montero-Camacho:2019jte}%
  \BibitemOpen
  \bibfield  {author} {\bibinfo {author} {\bibfnamefont {P.}~\bibnamefont
  {Montero-Camacho}}, \bibinfo {author} {\bibfnamefont {X.}~\bibnamefont
  {Fang}}, \bibinfo {author} {\bibfnamefont {G.}~\bibnamefont {Vasquez}},
  \bibinfo {author} {\bibfnamefont {M.}~\bibnamefont {Silva}}, \ and\ \bibinfo
  {author} {\bibfnamefont {C.~M.}\ \bibnamefont {Hirata}},\ }\href {\doibase
  10.1088/1475-7516/2019/08/031} {\bibfield  {journal} {\bibinfo  {journal}
  {JCAP}\ }\textbf {\bibinfo {volume} {1908}},\ \bibinfo {pages} {031}
  (\bibinfo {year} {2019})},\ \Eprint {http://arxiv.org/abs/1906.05950}
  {arXiv:1906.05950 [astro-ph.CO]} \BibitemShut {NoStop}%
\bibitem [{\citenamefont {Arzoumanian}\ \emph {et~al.}(2020)\citenamefont
  {Arzoumanian} \emph {et~al.}}]{Arzoumanian:2020vkk}%
  \BibitemOpen
  \bibfield  {author} {\bibinfo {author} {\bibfnamefont {Z.}~\bibnamefont
  {Arzoumanian}} \emph {et~al.} (\bibinfo {collaboration} {NANOGrav}),\
  }\href@noop {} {\  (\bibinfo {year} {2020})},\ \Eprint
  {http://arxiv.org/abs/2009.04496} {arXiv:2009.04496 [astro-ph.HE]}
  \BibitemShut {NoStop}%
\bibitem [{\citenamefont {Arzoumanian}\ \emph {et~al.}(2018)\citenamefont
  {Arzoumanian} \emph {et~al.}}]{Arzoumanian:2018saf}%
  \BibitemOpen
  \bibfield  {author} {\bibinfo {author} {\bibfnamefont {Z.}~\bibnamefont
  {Arzoumanian}} \emph {et~al.} (\bibinfo {collaboration} {NANOGRAV}),\ }\href
  {\doibase 10.3847/1538-4357/aabd3b} {\bibfield  {journal} {\bibinfo
  {journal} {Astrophys. J.}\ }\textbf {\bibinfo {volume} {859}},\ \bibinfo
  {pages} {47} (\bibinfo {year} {2018})},\ \Eprint
  {http://arxiv.org/abs/1801.02617} {arXiv:1801.02617 [astro-ph.HE]}
  \BibitemShut {NoStop}%
\bibitem [{\citenamefont {Shannon}\ \emph {et~al.}(2015)\citenamefont {Shannon}
  \emph {et~al.}}]{Shannon:2015ect}%
  \BibitemOpen
  \bibfield  {author} {\bibinfo {author} {\bibfnamefont {R.}~\bibnamefont
  {Shannon}} \emph {et~al.},\ }\href {\doibase 10.1126/science.aab1910}
  {\bibfield  {journal} {\bibinfo  {journal} {Science}\ }\textbf {\bibinfo
  {volume} {349}},\ \bibinfo {pages} {1522} (\bibinfo {year} {2015})},\ \Eprint
  {http://arxiv.org/abs/1509.07320} {arXiv:1509.07320 [astro-ph.CO]}
  \BibitemShut {NoStop}%
\bibitem [{\citenamefont {Sesana}\ \emph {et~al.}(2004)\citenamefont {Sesana},
  \citenamefont {Haardt}, \citenamefont {Madau},\ and\ \citenamefont
  {Volonteri}}]{Sesana:2004sp}%
  \BibitemOpen
  \bibfield  {author} {\bibinfo {author} {\bibfnamefont {A.}~\bibnamefont
  {Sesana}}, \bibinfo {author} {\bibfnamefont {F.}~\bibnamefont {Haardt}},
  \bibinfo {author} {\bibfnamefont {P.}~\bibnamefont {Madau}}, \ and\ \bibinfo
  {author} {\bibfnamefont {M.}~\bibnamefont {Volonteri}},\ }\href {\doibase
  10.1086/422185} {\bibfield  {journal} {\bibinfo  {journal} {Astrophys. J.}\
  }\textbf {\bibinfo {volume} {611}},\ \bibinfo {pages} {623} (\bibinfo {year}
  {2004})},\ \Eprint {http://arxiv.org/abs/astro-ph/0401543}
  {arXiv:astro-ph/0401543} \BibitemShut {NoStop}%
\bibitem [{\citenamefont {Burke-Spolaor}\ \emph {et~al.}(2019)\citenamefont
  {Burke-Spolaor} \emph {et~al.}}]{Burke-Spolaor:2018bvk}%
  \BibitemOpen
  \bibfield  {author} {\bibinfo {author} {\bibfnamefont {S.}~\bibnamefont
  {Burke-Spolaor}} \emph {et~al.},\ }\href {\doibase 10.1007/s00159-019-0115-7}
  {\bibfield  {journal} {\bibinfo  {journal} {Astron. Astrophys. Rev.}\
  }\textbf {\bibinfo {volume} {27}},\ \bibinfo {pages} {5} (\bibinfo {year}
  {2019})},\ \Eprint {http://arxiv.org/abs/1811.08826} {arXiv:1811.08826
  [astro-ph.HE]} \BibitemShut {NoStop}%
\bibitem [{\citenamefont {Blasi}\ \emph {et~al.}(2020)\citenamefont {Blasi},
  \citenamefont {Brdar},\ and\ \citenamefont {Schmitz}}]{Blasi:2020mfx}%
  \BibitemOpen
  \bibfield  {author} {\bibinfo {author} {\bibfnamefont {S.}~\bibnamefont
  {Blasi}}, \bibinfo {author} {\bibfnamefont {V.}~\bibnamefont {Brdar}}, \ and\
  \bibinfo {author} {\bibfnamefont {K.}~\bibnamefont {Schmitz}},\ }\href@noop
  {} {\  (\bibinfo {year} {2020})},\ \Eprint {http://arxiv.org/abs/2009.06607}
  {arXiv:2009.06607 [astro-ph.CO]} \BibitemShut {NoStop}%
\bibitem [{\citenamefont {Ellis}\ and\ \citenamefont
  {Lewicki}(2020)}]{Ellis:2020ena}%
  \BibitemOpen
  \bibfield  {author} {\bibinfo {author} {\bibfnamefont {J.}~\bibnamefont
  {Ellis}}\ and\ \bibinfo {author} {\bibfnamefont {M.}~\bibnamefont
  {Lewicki}},\ }\href@noop {} {\  (\bibinfo {year} {2020})},\ \Eprint
  {http://arxiv.org/abs/2009.06555} {arXiv:2009.06555 [astro-ph.CO]}
  \BibitemShut {NoStop}%
\bibitem [{\citenamefont {Buchmuller}\ \emph {et~al.}(2020)\citenamefont
  {Buchmuller}, \citenamefont {Domcke},\ and\ \citenamefont
  {Schmitz}}]{Buchmuller:2020lbh}%
  \BibitemOpen
  \bibfield  {author} {\bibinfo {author} {\bibfnamefont {W.}~\bibnamefont
  {Buchmuller}}, \bibinfo {author} {\bibfnamefont {V.}~\bibnamefont {Domcke}},
  \ and\ \bibinfo {author} {\bibfnamefont {K.}~\bibnamefont {Schmitz}},\
  }\href@noop {} {\  (\bibinfo {year} {2020})},\ \Eprint
  {http://arxiv.org/abs/2009.10649} {arXiv:2009.10649 [astro-ph.CO]}
  \BibitemShut {NoStop}%
\bibitem [{\citenamefont {Samanta}\ and\ \citenamefont
  {Datta}(2020)}]{Samanta:2020cdk}%
  \BibitemOpen
  \bibfield  {author} {\bibinfo {author} {\bibfnamefont {R.}~\bibnamefont
  {Samanta}}\ and\ \bibinfo {author} {\bibfnamefont {S.}~\bibnamefont
  {Datta}},\ }\href@noop {} {\  (\bibinfo {year} {2020})},\ \Eprint
  {http://arxiv.org/abs/2009.13452} {arXiv:2009.13452 [hep-ph]} \BibitemShut
  {NoStop}%
\bibitem [{\citenamefont {Bian}\ \emph {et~al.}(2020)\citenamefont {Bian},
  \citenamefont {Liu},\ and\ \citenamefont {Zhou}}]{Bian:2020bps}%
  \BibitemOpen
  \bibfield  {author} {\bibinfo {author} {\bibfnamefont {L.}~\bibnamefont
  {Bian}}, \bibinfo {author} {\bibfnamefont {J.}~\bibnamefont {Liu}}, \ and\
  \bibinfo {author} {\bibfnamefont {R.}~\bibnamefont {Zhou}},\ }\href@noop {}
  {\  (\bibinfo {year} {2020})},\ \Eprint {http://arxiv.org/abs/2009.13893}
  {arXiv:2009.13893 [astro-ph.CO]} \BibitemShut {NoStop}%
\bibitem [{\citenamefont {Nakai}\ \emph {et~al.}(2020)\citenamefont {Nakai},
  \citenamefont {Suzuki}, \citenamefont {Takahashi},\ and\ \citenamefont
  {Yamada}}]{Nakai:2020oit}%
  \BibitemOpen
  \bibfield  {author} {\bibinfo {author} {\bibfnamefont {Y.}~\bibnamefont
  {Nakai}}, \bibinfo {author} {\bibfnamefont {M.}~\bibnamefont {Suzuki}},
  \bibinfo {author} {\bibfnamefont {F.}~\bibnamefont {Takahashi}}, \ and\
  \bibinfo {author} {\bibfnamefont {M.}~\bibnamefont {Yamada}},\ }\href@noop {}
  {\  (\bibinfo {year} {2020})},\ \Eprint {http://arxiv.org/abs/2009.09754}
  {arXiv:2009.09754 [astro-ph.CO]} \BibitemShut {NoStop}%
\bibitem [{\citenamefont {Addazi}\ \emph {et~al.}(2020)\citenamefont {Addazi},
  \citenamefont {Cai}, \citenamefont {Gan}, \citenamefont {Marciano},\ and\
  \citenamefont {Zeng}}]{Addazi:2020zcj}%
  \BibitemOpen
  \bibfield  {author} {\bibinfo {author} {\bibfnamefont {A.}~\bibnamefont
  {Addazi}}, \bibinfo {author} {\bibfnamefont {Y.-F.}\ \bibnamefont {Cai}},
  \bibinfo {author} {\bibfnamefont {Q.}~\bibnamefont {Gan}}, \bibinfo {author}
  {\bibfnamefont {A.}~\bibnamefont {Marciano}}, \ and\ \bibinfo {author}
  {\bibfnamefont {K.}~\bibnamefont {Zeng}},\ }\href@noop {} {\  (\bibinfo
  {year} {2020})},\ \Eprint {http://arxiv.org/abs/2009.10327} {arXiv:2009.10327
  [hep-ph]} \BibitemShut {NoStop}%
\bibitem [{\citenamefont {Ratzinger}\ and\ \citenamefont
  {Schwaller}(2020)}]{Ratzinger:2020koh}%
  \BibitemOpen
  \bibfield  {author} {\bibinfo {author} {\bibfnamefont {W.}~\bibnamefont
  {Ratzinger}}\ and\ \bibinfo {author} {\bibfnamefont {P.}~\bibnamefont
  {Schwaller}},\ }\href@noop {} {\  (\bibinfo {year} {2020})},\ \Eprint
  {http://arxiv.org/abs/2009.11875} {arXiv:2009.11875 [astro-ph.CO]}
  \BibitemShut {NoStop}%
\bibitem [{\citenamefont {Li}\ \emph {et~al.}(2020)\citenamefont {Li},
  \citenamefont {Ye},\ and\ \citenamefont {Piao}}]{Li:2020cjj}%
  \BibitemOpen
  \bibfield  {author} {\bibinfo {author} {\bibfnamefont {H.-H.}\ \bibnamefont
  {Li}}, \bibinfo {author} {\bibfnamefont {G.}~\bibnamefont {Ye}}, \ and\
  \bibinfo {author} {\bibfnamefont {Y.-S.}\ \bibnamefont {Piao}},\ }\href@noop
  {} {\  (\bibinfo {year} {2020})},\ \Eprint {http://arxiv.org/abs/2009.14663}
  {arXiv:2009.14663 [astro-ph.CO]} \BibitemShut {NoStop}%
\bibitem [{\citenamefont {Vagnozzi}(2020)}]{Vagnozzi:2020gtf}%
  \BibitemOpen
  \bibfield  {author} {\bibinfo {author} {\bibfnamefont {S.}~\bibnamefont
  {Vagnozzi}},\ }\href@noop {} {\  (\bibinfo {year} {2020})},\ \Eprint
  {http://arxiv.org/abs/2009.13432} {arXiv:2009.13432 [astro-ph.CO]}
  \BibitemShut {NoStop}%
\bibitem [{\citenamefont {Vaskonen}\ and\ \citenamefont
  {Veerm\"ae}(2020)}]{Vaskonen:2020lbd}%
  \BibitemOpen
  \bibfield  {author} {\bibinfo {author} {\bibfnamefont {V.}~\bibnamefont
  {Vaskonen}}\ and\ \bibinfo {author} {\bibfnamefont {H.}~\bibnamefont
  {Veerm\"ae}},\ }\href@noop {} {\  (\bibinfo {year} {2020})},\ \Eprint
  {http://arxiv.org/abs/2009.07832} {arXiv:2009.07832 [astro-ph.CO]}
  \BibitemShut {NoStop}%
\bibitem [{\citenamefont {De~Luca}\ \emph
  {et~al.}(2020{\natexlab{a}})\citenamefont {De~Luca}, \citenamefont
  {Franciolini},\ and\ \citenamefont {Riotto}}]{DeLuca:2020agl}%
  \BibitemOpen
  \bibfield  {author} {\bibinfo {author} {\bibfnamefont {V.}~\bibnamefont
  {De~Luca}}, \bibinfo {author} {\bibfnamefont {G.}~\bibnamefont
  {Franciolini}}, \ and\ \bibinfo {author} {\bibfnamefont {A.}~\bibnamefont
  {Riotto}},\ }\href@noop {} {\  (\bibinfo {year} {2020}{\natexlab{a}})},\
  \Eprint {http://arxiv.org/abs/2009.08268} {arXiv:2009.08268 [astro-ph.CO]}
  \BibitemShut {NoStop}%
\bibitem [{\citenamefont {Kohri}\ and\ \citenamefont
  {Terada}(2020)}]{Kohri:2020qqd}%
  \BibitemOpen
  \bibfield  {author} {\bibinfo {author} {\bibfnamefont {K.}~\bibnamefont
  {Kohri}}\ and\ \bibinfo {author} {\bibfnamefont {T.}~\bibnamefont {Terada}},\
  }\href@noop {} {\  (\bibinfo {year} {2020})},\ \Eprint
  {http://arxiv.org/abs/2009.11853} {arXiv:2009.11853 [astro-ph.CO]}
  \BibitemShut {NoStop}%
\bibitem [{\citenamefont {De~Luca}\ \emph
  {et~al.}(2020{\natexlab{b}})\citenamefont {De~Luca}, \citenamefont
  {Franciolini},\ and\ \citenamefont {Riotto}}]{DeLuca:2020ioi}%
  \BibitemOpen
  \bibfield  {author} {\bibinfo {author} {\bibfnamefont {V.}~\bibnamefont
  {De~Luca}}, \bibinfo {author} {\bibfnamefont {G.}~\bibnamefont
  {Franciolini}}, \ and\ \bibinfo {author} {\bibfnamefont {A.}~\bibnamefont
  {Riotto}},\ }\href {\doibase 10.1016/j.physletb.2020.135550} {\bibfield
  {journal} {\bibinfo  {journal} {Phys. Lett. B}\ }\textbf {\bibinfo {volume}
  {807}},\ \bibinfo {pages} {135550} (\bibinfo {year} {2020}{\natexlab{b}})},\
  \Eprint {http://arxiv.org/abs/2001.04371} {arXiv:2001.04371 [astro-ph.CO]}
  \BibitemShut {NoStop}%
\bibitem [{\citenamefont {Wands}(1999)}]{Wands:1998yp}%
  \BibitemOpen
  \bibfield  {author} {\bibinfo {author} {\bibfnamefont {D.}~\bibnamefont
  {Wands}},\ }\href {\doibase 10.1103/PhysRevD.60.023507} {\bibfield  {journal}
  {\bibinfo  {journal} {Phys. Rev. D}\ }\textbf {\bibinfo {volume} {60}},\
  \bibinfo {pages} {023507} (\bibinfo {year} {1999})},\ \Eprint
  {http://arxiv.org/abs/gr-qc/9809062} {arXiv:gr-qc/9809062} \BibitemShut
  {NoStop}%
\bibitem [{\citenamefont {Leach}\ and\ \citenamefont
  {Liddle}(2001)}]{Leach:2000yw}%
  \BibitemOpen
  \bibfield  {author} {\bibinfo {author} {\bibfnamefont {S.~M.}\ \bibnamefont
  {Leach}}\ and\ \bibinfo {author} {\bibfnamefont {A.~R.}\ \bibnamefont
  {Liddle}},\ }\href {\doibase 10.1103/PhysRevD.63.043508} {\bibfield
  {journal} {\bibinfo  {journal} {Phys. Rev. D}\ }\textbf {\bibinfo {volume}
  {63}},\ \bibinfo {pages} {043508} (\bibinfo {year} {2001})},\ \Eprint
  {http://arxiv.org/abs/astro-ph/0010082} {arXiv:astro-ph/0010082} \BibitemShut
  {NoStop}%
\bibitem [{\citenamefont {Leach}\ \emph {et~al.}(2001)\citenamefont {Leach},
  \citenamefont {Sasaki}, \citenamefont {Wands},\ and\ \citenamefont
  {Liddle}}]{Leach:2001zf}%
  \BibitemOpen
  \bibfield  {author} {\bibinfo {author} {\bibfnamefont {S.~M.}\ \bibnamefont
  {Leach}}, \bibinfo {author} {\bibfnamefont {M.}~\bibnamefont {Sasaki}},
  \bibinfo {author} {\bibfnamefont {D.}~\bibnamefont {Wands}}, \ and\ \bibinfo
  {author} {\bibfnamefont {A.~R.}\ \bibnamefont {Liddle}},\ }\href {\doibase
  10.1103/PhysRevD.64.023512} {\bibfield  {journal} {\bibinfo  {journal} {Phys.
  Rev. D}\ }\textbf {\bibinfo {volume} {64}},\ \bibinfo {pages} {023512}
  (\bibinfo {year} {2001})},\ \Eprint {http://arxiv.org/abs/astro-ph/0101406}
  {arXiv:astro-ph/0101406} \BibitemShut {NoStop}%
\bibitem [{\citenamefont {Biagetti}\ \emph {et~al.}(2018)\citenamefont
  {Biagetti}, \citenamefont {Franciolini}, \citenamefont {Kehagias},\ and\
  \citenamefont {Riotto}}]{Biagetti:2018pjj}%
  \BibitemOpen
  \bibfield  {author} {\bibinfo {author} {\bibfnamefont {M.}~\bibnamefont
  {Biagetti}}, \bibinfo {author} {\bibfnamefont {G.}~\bibnamefont
  {Franciolini}}, \bibinfo {author} {\bibfnamefont {A.}~\bibnamefont
  {Kehagias}}, \ and\ \bibinfo {author} {\bibfnamefont {A.}~\bibnamefont
  {Riotto}},\ }\href {\doibase 10.1088/1475-7516/2018/07/032} {\bibfield
  {journal} {\bibinfo  {journal} {JCAP}\ }\textbf {\bibinfo {volume} {07}},\
  \bibinfo {pages} {032} (\bibinfo {year} {2018})},\ \Eprint
  {http://arxiv.org/abs/1804.07124} {arXiv:1804.07124 [astro-ph.CO]}
  \BibitemShut {NoStop}%
\bibitem [{\citenamefont {Niikura}\ \emph {et~al.}(2019)\citenamefont {Niikura}
  \emph {et~al.}}]{Niikura:2017zjd}%
  \BibitemOpen
  \bibfield  {author} {\bibinfo {author} {\bibfnamefont {H.}~\bibnamefont
  {Niikura}} \emph {et~al.},\ }\href {\doibase 10.1038/s41550-019-0723-1}
  {\bibfield  {journal} {\bibinfo  {journal} {Nat. Astron.}\ }\textbf {\bibinfo
  {volume} {3}},\ \bibinfo {pages} {524} (\bibinfo {year} {2019})},\ \Eprint
  {http://arxiv.org/abs/1701.02151} {arXiv:1701.02151 [astro-ph.CO]}
  \BibitemShut {NoStop}%
\bibitem [{\citenamefont {Sugiyama}\ \emph {et~al.}(2019)\citenamefont
  {Sugiyama}, \citenamefont {Kurita},\ and\ \citenamefont
  {Takada}}]{Sugiyama:2019dgt}%
  \BibitemOpen
  \bibfield  {author} {\bibinfo {author} {\bibfnamefont {S.}~\bibnamefont
  {Sugiyama}}, \bibinfo {author} {\bibfnamefont {T.}~\bibnamefont {Kurita}}, \
  and\ \bibinfo {author} {\bibfnamefont {M.}~\bibnamefont {Takada}},\
  }\href@noop {} {\  (\bibinfo {year} {2019})},\ \Eprint
  {http://arxiv.org/abs/1905.06066} {arXiv:1905.06066 [astro-ph.CO]}
  \BibitemShut {NoStop}%
\bibitem [{\citenamefont {Carr}\ \emph {et~al.}(2020)\citenamefont {Carr},
  \citenamefont {Kohri}, \citenamefont {Sendouda},\ and\ \citenamefont
  {Yokoyama}}]{Carr:2020gox}%
  \BibitemOpen
  \bibfield  {author} {\bibinfo {author} {\bibfnamefont {B.}~\bibnamefont
  {Carr}}, \bibinfo {author} {\bibfnamefont {K.}~\bibnamefont {Kohri}},
  \bibinfo {author} {\bibfnamefont {Y.}~\bibnamefont {Sendouda}}, \ and\
  \bibinfo {author} {\bibfnamefont {J.}~\bibnamefont {Yokoyama}},\ }\href@noop
  {} {\  (\bibinfo {year} {2020})},\ \Eprint {http://arxiv.org/abs/2002.12778}
  {arXiv:2002.12778 [astro-ph.CO]} \BibitemShut {NoStop}%
\bibitem [{\citenamefont {Arbey}\ \emph {et~al.}(2020)\citenamefont {Arbey},
  \citenamefont {Auffinger},\ and\ \citenamefont {Silk}}]{Arbey:2019vqx}%
  \BibitemOpen
  \bibfield  {author} {\bibinfo {author} {\bibfnamefont {A.}~\bibnamefont
  {Arbey}}, \bibinfo {author} {\bibfnamefont {J.}~\bibnamefont {Auffinger}}, \
  and\ \bibinfo {author} {\bibfnamefont {J.}~\bibnamefont {Silk}},\ }\href
  {\doibase 10.1103/PhysRevD.101.023010} {\bibfield  {journal} {\bibinfo
  {journal} {Phys. Rev.}\ }\textbf {\bibinfo {volume} {D101}},\ \bibinfo
  {pages} {023010} (\bibinfo {year} {2020})},\ \Eprint
  {http://arxiv.org/abs/1906.04750} {arXiv:1906.04750 [astro-ph.CO]}
  \BibitemShut {NoStop}%
\bibitem [{\citenamefont {Dasgupta}\ \emph {et~al.}(2019)\citenamefont
  {Dasgupta}, \citenamefont {Laha},\ and\ \citenamefont
  {Ray}}]{Dasgupta:2019cae}%
  \BibitemOpen
  \bibfield  {author} {\bibinfo {author} {\bibfnamefont {B.}~\bibnamefont
  {Dasgupta}}, \bibinfo {author} {\bibfnamefont {R.}~\bibnamefont {Laha}}, \
  and\ \bibinfo {author} {\bibfnamefont {A.}~\bibnamefont {Ray}},\ }\href@noop
  {} {\  (\bibinfo {year} {2019})},\ \Eprint {http://arxiv.org/abs/1912.01014}
  {arXiv:1912.01014 [hep-ph]} \BibitemShut {NoStop}%
\bibitem [{\citenamefont {Laha}(2019)}]{Laha:2019ssq}%
  \BibitemOpen
  \bibfield  {author} {\bibinfo {author} {\bibfnamefont {R.}~\bibnamefont
  {Laha}},\ }\href {\doibase 10.1103/PhysRevLett.123.251101} {\bibfield
  {journal} {\bibinfo  {journal} {Phys. Rev. Lett.}\ }\textbf {\bibinfo
  {volume} {123}},\ \bibinfo {pages} {251101} (\bibinfo {year} {2019})},\
  \Eprint {http://arxiv.org/abs/1906.09994} {arXiv:1906.09994 [astro-ph.HE]}
  \BibitemShut {NoStop}%
\bibitem [{\citenamefont {DeRocco}\ and\ \citenamefont
  {Graham}(2019)}]{DeRocco:2019fjq}%
  \BibitemOpen
  \bibfield  {author} {\bibinfo {author} {\bibfnamefont {W.}~\bibnamefont
  {DeRocco}}\ and\ \bibinfo {author} {\bibfnamefont {P.~W.}\ \bibnamefont
  {Graham}},\ }\href {\doibase 10.1103/PhysRevLett.123.251102} {\bibfield
  {journal} {\bibinfo  {journal} {Phys. Rev. Lett.}\ }\textbf {\bibinfo
  {volume} {123}},\ \bibinfo {pages} {251102} (\bibinfo {year} {2019})},\
  \Eprint {http://arxiv.org/abs/1906.07740} {arXiv:1906.07740 [astro-ph.CO]}
  \BibitemShut {NoStop}%
\bibitem [{\citenamefont {Laha}\ \emph
  {et~al.}(2020{\natexlab{a}})\citenamefont {Laha}, \citenamefont {Lu},\ and\
  \citenamefont {Takhistov}}]{Laha:2020vhg}%
  \BibitemOpen
  \bibfield  {author} {\bibinfo {author} {\bibfnamefont {R.}~\bibnamefont
  {Laha}}, \bibinfo {author} {\bibfnamefont {P.}~\bibnamefont {Lu}}, \ and\
  \bibinfo {author} {\bibfnamefont {V.}~\bibnamefont {Takhistov}},\ }\href@noop
  {} {\  (\bibinfo {year} {2020}{\natexlab{a}})},\ \Eprint
  {http://arxiv.org/abs/2009.11837} {arXiv:2009.11837 [astro-ph.CO]}
  \BibitemShut {NoStop}%
\bibitem [{\citenamefont {Laha}\ \emph
  {et~al.}(2020{\natexlab{b}})\citenamefont {Laha}, \citenamefont {Mu\~noz},\
  and\ \citenamefont {Slatyer}}]{Laha:2020ivk}%
  \BibitemOpen
  \bibfield  {author} {\bibinfo {author} {\bibfnamefont {R.}~\bibnamefont
  {Laha}}, \bibinfo {author} {\bibfnamefont {J.~B.}\ \bibnamefont {Mu\~noz}}, \
  and\ \bibinfo {author} {\bibfnamefont {T.~R.}\ \bibnamefont {Slatyer}},\
  }\href {\doibase 10.1103/PhysRevD.101.123514} {\bibfield  {journal} {\bibinfo
   {journal} {Phys. Rev. D}\ }\textbf {\bibinfo {volume} {101}},\ \bibinfo
  {pages} {123514} (\bibinfo {year} {2020}{\natexlab{b}})},\ \Eprint
  {http://arxiv.org/abs/2004.00627} {arXiv:2004.00627 [astro-ph.CO]}
  \BibitemShut {NoStop}%
\bibitem [{\citenamefont {Griest}\ \emph {et~al.}(2014)\citenamefont {Griest},
  \citenamefont {Cieplak},\ and\ \citenamefont {Lehner}}]{Griest:2013aaa}%
  \BibitemOpen
  \bibfield  {author} {\bibinfo {author} {\bibfnamefont {K.}~\bibnamefont
  {Griest}}, \bibinfo {author} {\bibfnamefont {A.~M.}\ \bibnamefont {Cieplak}},
  \ and\ \bibinfo {author} {\bibfnamefont {M.~J.}\ \bibnamefont {Lehner}},\
  }\href {\doibase 10.1088/0004-637X/786/2/158} {\bibfield  {journal} {\bibinfo
   {journal} {Astrophys. J.}\ }\textbf {\bibinfo {volume} {786}},\ \bibinfo
  {pages} {158} (\bibinfo {year} {2014})},\ \Eprint
  {http://arxiv.org/abs/1307.5798} {arXiv:1307.5798 [astro-ph.CO]} \BibitemShut
  {NoStop}%
\bibitem [{\citenamefont {Tisserand}\ \emph {et~al.}(2007)\citenamefont
  {Tisserand} \emph {et~al.}}]{Tisserand:2006zx}%
  \BibitemOpen
  \bibfield  {author} {\bibinfo {author} {\bibfnamefont {P.}~\bibnamefont
  {Tisserand}} \emph {et~al.} (\bibinfo {collaboration} {EROS-2}),\ }\href
  {\doibase 10.1051/0004-6361:20066017} {\bibfield  {journal} {\bibinfo
  {journal} {Astron. Astrophys.}\ }\textbf {\bibinfo {volume} {469}},\ \bibinfo
  {pages} {387} (\bibinfo {year} {2007})},\ \Eprint
  {http://arxiv.org/abs/astro-ph/0607207} {arXiv:astro-ph/0607207 [astro-ph]}
  \BibitemShut {NoStop}%
\bibitem [{\citenamefont {Ali-Haïmoud}\ and\ \citenamefont
  {Kamionkowski}(2017)}]{Ali-Haimoud:2016mbv}%
  \BibitemOpen
  \bibfield  {author} {\bibinfo {author} {\bibfnamefont {Y.}~\bibnamefont
  {Ali-Haïmoud}}\ and\ \bibinfo {author} {\bibfnamefont {M.}~\bibnamefont
  {Kamionkowski}},\ }\href {\doibase 10.1103/PhysRevD.95.043534} {\bibfield
  {journal} {\bibinfo  {journal} {Phys. Rev.}\ }\textbf {\bibinfo {volume}
  {D95}},\ \bibinfo {pages} {043534} (\bibinfo {year} {2017})},\ \Eprint
  {http://arxiv.org/abs/1612.05644} {arXiv:1612.05644 [astro-ph.CO]}
  \BibitemShut {NoStop}%
\bibitem [{\citenamefont {Poulin}\ \emph {et~al.}(2017)\citenamefont {Poulin},
  \citenamefont {Serpico}, \citenamefont {Calore}, \citenamefont {Clesse},\
  and\ \citenamefont {Kohri}}]{Poulin:2017bwe}%
  \BibitemOpen
  \bibfield  {author} {\bibinfo {author} {\bibfnamefont {V.}~\bibnamefont
  {Poulin}}, \bibinfo {author} {\bibfnamefont {P.~D.}\ \bibnamefont {Serpico}},
  \bibinfo {author} {\bibfnamefont {F.}~\bibnamefont {Calore}}, \bibinfo
  {author} {\bibfnamefont {S.}~\bibnamefont {Clesse}}, \ and\ \bibinfo {author}
  {\bibfnamefont {K.}~\bibnamefont {Kohri}},\ }\href {\doibase
  10.1103/PhysRevD.96.083524} {\bibfield  {journal} {\bibinfo  {journal} {Phys.
  Rev.}\ }\textbf {\bibinfo {volume} {D96}},\ \bibinfo {pages} {083524}
  (\bibinfo {year} {2017})},\ \Eprint {http://arxiv.org/abs/1707.04206}
  {arXiv:1707.04206 [astro-ph.CO]} \BibitemShut {NoStop}%
\bibitem [{\citenamefont {Poulter}\ \emph {et~al.}(2019)\citenamefont
  {Poulter}, \citenamefont {Ali-Haïmoud}, \citenamefont {Hamann},
  \citenamefont {White},\ and\ \citenamefont {Williams}}]{Poulter:2019ooo}%
  \BibitemOpen
  \bibfield  {author} {\bibinfo {author} {\bibfnamefont {H.}~\bibnamefont
  {Poulter}}, \bibinfo {author} {\bibfnamefont {Y.}~\bibnamefont
  {Ali-Haïmoud}}, \bibinfo {author} {\bibfnamefont {J.}~\bibnamefont
  {Hamann}}, \bibinfo {author} {\bibfnamefont {M.}~\bibnamefont {White}}, \
  and\ \bibinfo {author} {\bibfnamefont {A.~G.}\ \bibnamefont {Williams}},\
  }\href@noop {} {\  (\bibinfo {year} {2019})},\ \Eprint
  {http://arxiv.org/abs/1907.06485} {arXiv:1907.06485 [astro-ph.CO]}
  \BibitemShut {NoStop}%
\bibitem [{\citenamefont {Lu}\ \emph {et~al.}(2020)\citenamefont {Lu},
  \citenamefont {Takhistov}, \citenamefont {Gelmini}, \citenamefont {Hayashi},
  \citenamefont {Inoue},\ and\ \citenamefont {Kusenko}}]{Lu:2020bmd}%
  \BibitemOpen
  \bibfield  {author} {\bibinfo {author} {\bibfnamefont {P.}~\bibnamefont
  {Lu}}, \bibinfo {author} {\bibfnamefont {V.}~\bibnamefont {Takhistov}},
  \bibinfo {author} {\bibfnamefont {G.~B.}\ \bibnamefont {Gelmini}}, \bibinfo
  {author} {\bibfnamefont {K.}~\bibnamefont {Hayashi}}, \bibinfo {author}
  {\bibfnamefont {Y.}~\bibnamefont {Inoue}}, \ and\ \bibinfo {author}
  {\bibfnamefont {A.}~\bibnamefont {Kusenko}},\ }\href@noop {} {\  (\bibinfo
  {year} {2020})},\ \Eprint {http://arxiv.org/abs/2007.02213} {arXiv:2007.02213
  [astro-ph.CO]} \BibitemShut {NoStop}%
\bibitem [{\citenamefont {Choptuik}(1993)}]{Choptuik:1993}%
  \BibitemOpen
  \bibfield  {author} {\bibinfo {author} {\bibfnamefont {M.~W.}\ \bibnamefont
  {Choptuik}},\ }\href {\doibase 10.1103/PhysRevLett.70.9} {\bibfield
  {journal} {\bibinfo  {journal} {Phys. Rev. Lett.}\ }\textbf {\bibinfo
  {volume} {70}},\ \bibinfo {pages} {9} (\bibinfo {year} {1993})}\BibitemShut
  {NoStop}%
\bibitem [{\citenamefont {Evans}\ and\ \citenamefont
  {Coleman}(1994)}]{Evans:1994pj}%
  \BibitemOpen
  \bibfield  {author} {\bibinfo {author} {\bibfnamefont {C.~R.}\ \bibnamefont
  {Evans}}\ and\ \bibinfo {author} {\bibfnamefont {J.~S.}\ \bibnamefont
  {Coleman}},\ }\href {\doibase 10.1103/PhysRevLett.72.1782} {\bibfield
  {journal} {\bibinfo  {journal} {Phys. Rev. Lett.}\ }\textbf {\bibinfo
  {volume} {72}},\ \bibinfo {pages} {1782} (\bibinfo {year} {1994})},\ \Eprint
  {http://arxiv.org/abs/gr-qc/9402041} {arXiv:gr-qc/9402041} \BibitemShut
  {NoStop}%
\bibitem [{\citenamefont {Niemeyer}\ and\ \citenamefont
  {Jedamzik}(1998)}]{Niemeyer:1997mt}%
  \BibitemOpen
  \bibfield  {author} {\bibinfo {author} {\bibfnamefont {J.~C.}\ \bibnamefont
  {Niemeyer}}\ and\ \bibinfo {author} {\bibfnamefont {K.}~\bibnamefont
  {Jedamzik}},\ }\href {\doibase 10.1103/PhysRevLett.80.5481} {\bibfield
  {journal} {\bibinfo  {journal} {Phys. Rev. Lett.}\ }\textbf {\bibinfo
  {volume} {80}},\ \bibinfo {pages} {5481} (\bibinfo {year} {1998})},\ \Eprint
  {http://arxiv.org/abs/astro-ph/9709072} {arXiv:astro-ph/9709072} \BibitemShut
  {NoStop}%
\bibitem [{\citenamefont {Niemeyer}\ and\ \citenamefont
  {Jedamzik}(1999)}]{Niemeyer:1999ak}%
  \BibitemOpen
  \bibfield  {author} {\bibinfo {author} {\bibfnamefont {J.~C.}\ \bibnamefont
  {Niemeyer}}\ and\ \bibinfo {author} {\bibfnamefont {K.}~\bibnamefont
  {Jedamzik}},\ }\href {\doibase 10.1103/PhysRevD.59.124013} {\bibfield
  {journal} {\bibinfo  {journal} {Phys. Rev. D}\ }\textbf {\bibinfo {volume}
  {59}},\ \bibinfo {pages} {124013} (\bibinfo {year} {1999})},\ \Eprint
  {http://arxiv.org/abs/astro-ph/9901292} {arXiv:astro-ph/9901292} \BibitemShut
  {NoStop}%
\bibitem [{\citenamefont {Koike}\ \emph {et~al.}(1995)\citenamefont {Koike},
  \citenamefont {Hara},\ and\ \citenamefont {Adachi}}]{Koike:1995jm}%
  \BibitemOpen
  \bibfield  {author} {\bibinfo {author} {\bibfnamefont {T.}~\bibnamefont
  {Koike}}, \bibinfo {author} {\bibfnamefont {T.}~\bibnamefont {Hara}}, \ and\
  \bibinfo {author} {\bibfnamefont {S.}~\bibnamefont {Adachi}},\ }\href
  {\doibase 10.1103/PhysRevLett.74.5170} {\bibfield  {journal} {\bibinfo
  {journal} {Phys. Rev. Lett.}\ }\textbf {\bibinfo {volume} {74}},\ \bibinfo
  {pages} {5170} (\bibinfo {year} {1995})},\ \Eprint
  {http://arxiv.org/abs/gr-qc/9503007} {arXiv:gr-qc/9503007} \BibitemShut
  {NoStop}%
\bibitem [{\citenamefont {Byrnes}\ \emph {et~al.}(2018)\citenamefont {Byrnes},
  \citenamefont {Hindmarsh}, \citenamefont {Young},\ and\ \citenamefont
  {Hawkins}}]{Byrnes:2018clq}%
  \BibitemOpen
  \bibfield  {author} {\bibinfo {author} {\bibfnamefont {C.~T.}\ \bibnamefont
  {Byrnes}}, \bibinfo {author} {\bibfnamefont {M.}~\bibnamefont {Hindmarsh}},
  \bibinfo {author} {\bibfnamefont {S.}~\bibnamefont {Young}}, \ and\ \bibinfo
  {author} {\bibfnamefont {M.~R.~S.}\ \bibnamefont {Hawkins}},\ }\href
  {\doibase 10.1088/1475-7516/2018/08/041} {\bibfield  {journal} {\bibinfo
  {journal} {JCAP}\ }\textbf {\bibinfo {volume} {08}},\ \bibinfo {pages} {041}
  (\bibinfo {year} {2018})},\ \Eprint {http://arxiv.org/abs/1801.06138}
  {arXiv:1801.06138 [astro-ph.CO]} \BibitemShut {NoStop}%
\bibitem [{\citenamefont {Cai}\ \emph {et~al.}(2019)\citenamefont {Cai},
  \citenamefont {Pi},\ and\ \citenamefont {Sasaki}}]{Cai:2018dig}%
  \BibitemOpen
  \bibfield  {author} {\bibinfo {author} {\bibfnamefont {R.-g.}\ \bibnamefont
  {Cai}}, \bibinfo {author} {\bibfnamefont {S.}~\bibnamefont {Pi}}, \ and\
  \bibinfo {author} {\bibfnamefont {M.}~\bibnamefont {Sasaki}},\ }\href
  {\doibase 10.1103/PhysRevLett.122.201101} {\bibfield  {journal} {\bibinfo
  {journal} {Phys. Rev. Lett.}\ }\textbf {\bibinfo {volume} {122}},\ \bibinfo
  {pages} {201101} (\bibinfo {year} {2019})},\ \Eprint
  {http://arxiv.org/abs/1810.11000} {arXiv:1810.11000 [astro-ph.CO]}
  \BibitemShut {NoStop}%
\bibitem [{\citenamefont {Ananda}\ \emph {et~al.}(2007)\citenamefont {Ananda},
  \citenamefont {Clarkson},\ and\ \citenamefont {Wands}}]{Ananda:2006af}%
  \BibitemOpen
  \bibfield  {author} {\bibinfo {author} {\bibfnamefont {K.~N.}\ \bibnamefont
  {Ananda}}, \bibinfo {author} {\bibfnamefont {C.}~\bibnamefont {Clarkson}}, \
  and\ \bibinfo {author} {\bibfnamefont {D.}~\bibnamefont {Wands}},\ }\href
  {\doibase 10.1103/PhysRevD.75.123518} {\bibfield  {journal} {\bibinfo
  {journal} {Phys. Rev. D}\ }\textbf {\bibinfo {volume} {75}},\ \bibinfo
  {pages} {123518} (\bibinfo {year} {2007})},\ \Eprint
  {http://arxiv.org/abs/gr-qc/0612013} {arXiv:gr-qc/0612013} \BibitemShut
  {NoStop}%
\bibitem [{\citenamefont {Kohri}\ and\ \citenamefont
  {Terada}(2018)}]{Kohri:2018awv}%
  \BibitemOpen
  \bibfield  {author} {\bibinfo {author} {\bibfnamefont {K.}~\bibnamefont
  {Kohri}}\ and\ \bibinfo {author} {\bibfnamefont {T.}~\bibnamefont {Terada}},\
  }\href {\doibase 10.1103/PhysRevD.97.123532} {\bibfield  {journal} {\bibinfo
  {journal} {Phys. Rev. D}\ }\textbf {\bibinfo {volume} {97}},\ \bibinfo
  {pages} {123532} (\bibinfo {year} {2018})},\ \Eprint
  {http://arxiv.org/abs/1804.08577} {arXiv:1804.08577 [gr-qc]} \BibitemShut
  {NoStop}%
\bibitem [{\citenamefont {Espinosa}\ \emph {et~al.}(2018)\citenamefont
  {Espinosa}, \citenamefont {Racco},\ and\ \citenamefont
  {Riotto}}]{Espinosa:2018eve}%
  \BibitemOpen
  \bibfield  {author} {\bibinfo {author} {\bibfnamefont {J.~R.}\ \bibnamefont
  {Espinosa}}, \bibinfo {author} {\bibfnamefont {D.}~\bibnamefont {Racco}}, \
  and\ \bibinfo {author} {\bibfnamefont {A.}~\bibnamefont {Riotto}},\ }\href
  {\doibase 10.1088/1475-7516/2018/09/012} {\bibfield  {journal} {\bibinfo
  {journal} {JCAP}\ }\textbf {\bibinfo {volume} {09}},\ \bibinfo {pages} {012}
  (\bibinfo {year} {2018})},\ \Eprint {http://arxiv.org/abs/1804.07732}
  {arXiv:1804.07732 [hep-ph]} \BibitemShut {NoStop}%
\bibitem [{\citenamefont {Hajkarim}\ and\ \citenamefont
  {Schaffner-Bielich}(2020)}]{Hajkarim:2019nbx}%
  \BibitemOpen
  \bibfield  {author} {\bibinfo {author} {\bibfnamefont {F.}~\bibnamefont
  {Hajkarim}}\ and\ \bibinfo {author} {\bibfnamefont {J.}~\bibnamefont
  {Schaffner-Bielich}},\ }\href {\doibase 10.1103/PhysRevD.101.043522}
  {\bibfield  {journal} {\bibinfo  {journal} {Phys. Rev. D}\ }\textbf {\bibinfo
  {volume} {101}},\ \bibinfo {pages} {043522} (\bibinfo {year} {2020})},\
  \Eprint {http://arxiv.org/abs/1910.12357} {arXiv:1910.12357 [hep-ph]}
  \BibitemShut {NoStop}%
\bibitem [{\citenamefont {Oguri}\ \emph {et~al.}(2018)\citenamefont {Oguri},
  \citenamefont {Diego}, \citenamefont {Kaiser}, \citenamefont {Kelly},\ and\
  \citenamefont {Broadhurst}}]{Oguri:2017ock}%
  \BibitemOpen
  \bibfield  {author} {\bibinfo {author} {\bibfnamefont {M.}~\bibnamefont
  {Oguri}}, \bibinfo {author} {\bibfnamefont {J.~M.}\ \bibnamefont {Diego}},
  \bibinfo {author} {\bibfnamefont {N.}~\bibnamefont {Kaiser}}, \bibinfo
  {author} {\bibfnamefont {P.~L.}\ \bibnamefont {Kelly}}, \ and\ \bibinfo
  {author} {\bibfnamefont {T.}~\bibnamefont {Broadhurst}},\ }\href {\doibase
  10.1103/PhysRevD.97.023518} {\bibfield  {journal} {\bibinfo  {journal} {Phys.
  Rev. D}\ }\textbf {\bibinfo {volume} {97}},\ \bibinfo {pages} {023518}
  (\bibinfo {year} {2018})},\ \Eprint {http://arxiv.org/abs/1710.00148}
  {arXiv:1710.00148 [astro-ph.CO]} \BibitemShut {NoStop}%
\bibitem [{\citenamefont {Ali-Ha\"\i{}moud}\ \emph {et~al.}(2017)\citenamefont
  {Ali-Ha\"\i{}moud}, \citenamefont {Kovetz},\ and\ \citenamefont
  {Kamionkowski}}]{Ali-Haimoud:2017rtz}%
  \BibitemOpen
  \bibfield  {author} {\bibinfo {author} {\bibfnamefont {Y.}~\bibnamefont
  {Ali-Ha\"\i{}moud}}, \bibinfo {author} {\bibfnamefont {E.~D.}\ \bibnamefont
  {Kovetz}}, \ and\ \bibinfo {author} {\bibfnamefont {M.}~\bibnamefont
  {Kamionkowski}},\ }\href {\doibase 10.1103/PhysRevD.96.123523} {\bibfield
  {journal} {\bibinfo  {journal} {Phys. Rev. D}\ }\textbf {\bibinfo {volume}
  {96}},\ \bibinfo {pages} {123523} (\bibinfo {year} {2017})},\ \Eprint
  {http://arxiv.org/abs/1709.06576} {arXiv:1709.06576 [astro-ph.CO]}
  \BibitemShut {NoStop}%
\end{thebibliography}%
 
\end{document}